\documentclass{article}

\usepackage[preprint]{neurips_2025}
\usepackage{svg}

\usepackage[utf8]{inputenc} 
\usepackage[T1]{fontenc}    
\usepackage{hyperref}       
\usepackage{url}            
\usepackage{booktabs}       
\usepackage{amsfonts}       

\usepackage{caption}
\usepackage{graphicx}     
\usepackage{subcaption}
\usepackage{array}
\usepackage[export]{adjustbox}

\usepackage{nicefrac}       
\usepackage{microtype}      
\usepackage{amsmath}

\title{VEIGAR: View-consistent Explicit Inpainting and Geometry Alignment for 3D object Removal}

\author{%
  Pham Khai Nguyen Do, Bao Nguyen Tran, Nam Nguyen, Duc Dung Nguyen\thanks{Correspondence} \\
  AITech Lab, Computer Science and Engineering Faculty\\
  Ho Chi Minh City University of Technology, VNUHCM\\
  \texttt{\{nddung\}@hcmut.edu.vn} \\
}

\begin{document}

\maketitle

\begin{abstract}

Recent advances in Novel View Synthesis (NVS) and 3D generation have significantly improved editing tasks, with a primary emphasis on maintaining cross-view consistency throughout the generative process. Contemporary methods typically address this challenge using a dual-strategy framework: performing consistent 2D inpainting across all views guided by embedded priors either explicitly in pixel space or implicitly in latent space; and conducting 3D reconstruction with additional consistency guidance. Previous strategies, in particular, often require an initial 3D reconstruction phase to establish geometric structure, introducing considerable computational overhead. Even with the added cost, the resulting reconstruction quality often remains suboptimal. In this paper, we present VEIGAR, a computationally efficient framework that outperforms existing methods without relying on an initial reconstruction phase. VEIGAR leverages a lightweight foundation model to reliably align priors explicitly in the pixel space. In addition, we introduce a novel supervision strategy based on scale-invariant depth loss, which removes the need for traditional scale-and-shift operations in monocular depth regularization. Through extensive experimentation, VEIGAR establishes a new state-of-the-art benchmark in reconstruction quality and cross-view consistency, while achieving a threefold reduction in training time compared to the fastest existing method, highlighting its superior balance of efficiency and effectiveness. 


\end{abstract}

\begin{figure}
    \centering
    \includegraphics[width=0.85\linewidth]{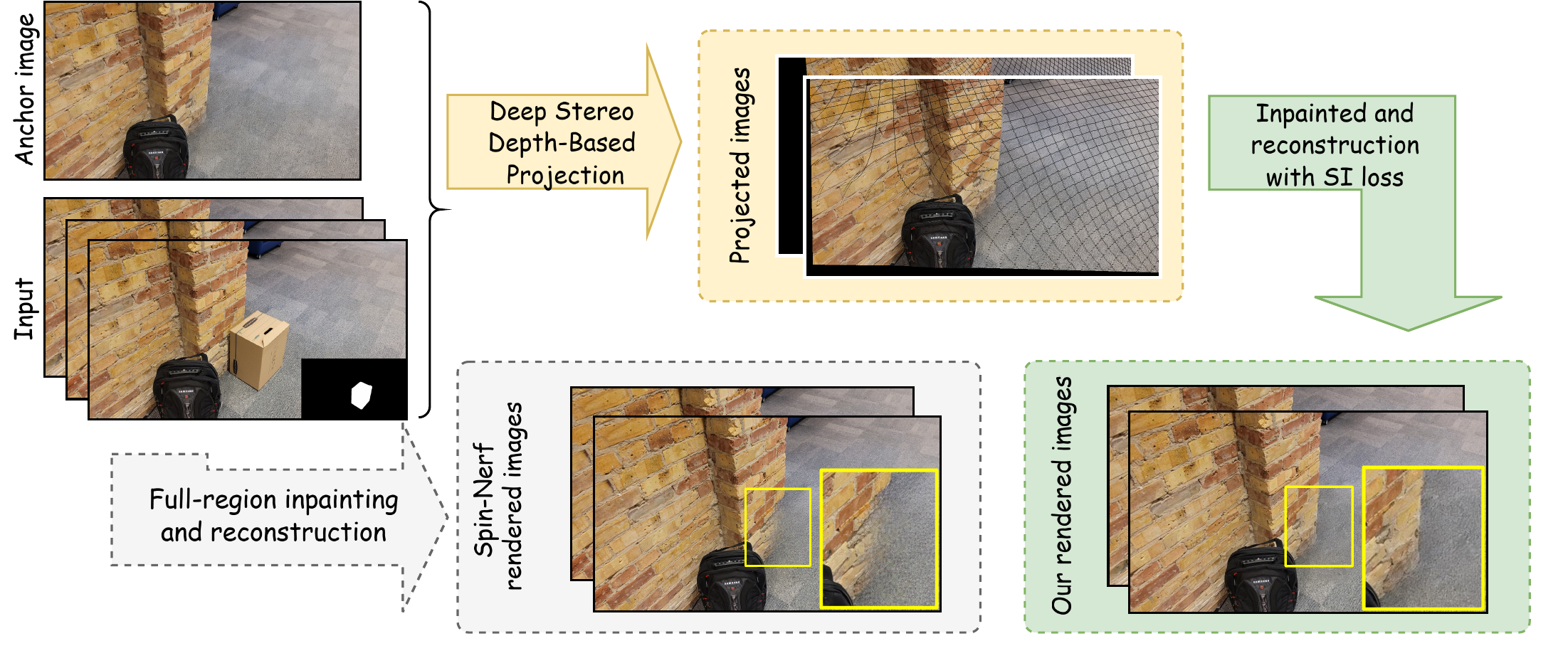}
    \caption{An illustration of the VEIGAR inpainting and reconstruction pipeline. VEIGAR inpaints a single anchor view and propagates edits to other views using deep stereo depth-based projection, followed by completion and reconstruction guided by a scale-invariant depth loss. Compared to prior methods like SPIn-NeRF, VEIGAR produces sharper and more view-consistent results.}
    \label{fig:intro}
\end{figure}

\section{Introduction}
Novel view synthesis (NVS) has become a key task in computer vision, crucial for 3D scene reconstruction, immersive content creation, and VR/AR applications. Advances like Neural Radiance Fields (NeRF)~\cite{mildenhall2020nerf} and 3D Gaussian Splatting (3DGS)~\cite{kerbl3Dgaussians} have significantly advanced in photorealistic rendering from sparse multi-view observations. However, real-world applications such as scene editing, object manipulation, and view modification require more than faithful reconstruction. In particular, achieving object removal while preserving photorealism and geometric consistency across views remains challenging, as standard inpainting methods often introduce artifacts and cross-view inconsistencies due to a lack of 3D awareness.

To address the challenge of cross-view inconsistency in novel view synthesis and 3D generation, prior works have commonly adopted one or both of two main strategies. The first involves applying inpainting techniques to enforce multiview consistency across all views, while the second incorporates additional guidance during the 3D reconstruction stage to mitigate conflicts arising from matching errors. The first strategy builds upon advances in 2D inpainting methods~\cite{pathak2016context, iizuka2017globally, yu2018generative, liu2018image, suvorov2022resolution}, which have been extended to the multiview setting by incorporating cross-view attention mechanisms~\cite{huang2025_2502.11801, shi2024crossview, gaussctrl2024, chen2023gaussianeditor, salimi2025geometry, cerkezi2023multi, cao2024mvinpainter} or reference-based method~\cite{hu2024innout, wang2024gscream, lu2024view}. However, these improvements are often constrained to the semantic level and fail to ensure precise pixel-wise alignment, which is crucial for accurate 3D reconstruction. Moreover, reference-based methods typically rely on accurate depth maps, the estimation of which demands a computationally expensive initial training phase. On the other hand, the second strategy focuses on guiding the reconstruction process through global scene optimization. This is often achieved by leveraging Score Distillation Sampling (SDS)~\cite{alldieck2024score, bahmani2024fourdfy, yang2024semantic, poole2023dreamfusion, Liang_2024_CVPR}, allowing diffusion models to distill and globally update the scene representation. While this enables consistent changes across the entire scene, it often comes at the cost of reduced visual fidelity and significant computational overhead. Another common form of guidance during reconstruction is geometry supervision~\cite{spinnerf, wang2024gscream, liu2024infusioninpainting3dgaussians}, typically provided through depth estimation~\cite{zoedepth, yang2024depth, chung2023depth, yang2024depthanything}. This helps maintain geometric coherence across views. Nevertheless, such approaches suffer from inherent scale ambiguity in monocular depth predictions, necessitating a scale-and-shift normalization step to align reconstructions properly.

To this end, we propose VEIGAR (View-consistent Explicit Inpainting and Geometry Alignment for Object Removal), a unified and efficient framework designed for 3D object removal within the Gaussian Splatting paradigm. An overview of the proposed pipeline is illustrated in Fig.~\ref{fig:intro}. VEIGAR enhances cross-view consistency by aligning features explicitly at the pixel level across all views, bypassing the need for latent-space alignment. To mitigate the computational overhead of an initial reconstruction phase, VEIGAR utilizes a stereo-pretrained depth estimator to provide accurate geometric guidance for projection. During reconstruction, we introduce a novel supervision mechanism based on \textit{scale-invariant depth loss}, which eliminates the conventional scale-and-shift step required in monocular depth regularization. This loss enforces structural consistency throughout the entire reconstruction process, rather than limiting it to early-stage geometry formation. Extensive experiments demonstrate that VEIGAR surpasses existing baselines in visual fidelity, geometric consistency, and training efficiency. Project page: \href{https://veigar3d.github.io/VEIGAR}{https://veigar3d.github.io/VEIGAR}

In summary, our contributions are threefold:
(1) We present VEIGAR, a computationally efficient framework that surpasses existing methods without relying on an initial reconstruction step;
(2) VEIGAR employs a pre-trained model to explicitly and reliably align global priors at the pixel level, enhancing cross-view consistency;
(3) We introduce a novel supervision strategy using scale-invariant depth loss that removes the dependency on traditional scale-and-shift operations in monocular depth.

\section{Related Works}
\paragraph{3D Scene Editing via Image Inpainting}
Diffusion models have demonstrated remarkable capabilities in high-fidelity image generation and editing tasks \cite{sohl2015deep, ho2020denoising, nichol2021improved, song2021scorebased}, particularly in the context of image inpainting \cite{lugmayr2022repaint, saharia2022palette, avrahami2022blended}. Recent efforts have explored extending these methods to 3D scene editing by leveraging the generative prior of diffusion models to synthesize view-consistent images from sparse observations \cite{watson2022novelview, liu2023zero123, zhang2023diffusion, wang2023zero123plusplus}. However, lifting 2D inpainting to 3D introduces the critical challenge of maintaining multiview consistency, which is essential for downstream tasks like novel view synthesis and 3D reconstruction. Several works address this by incorporating reference-guided or temporal inpainting mechanisms \cite{qian2022make, avrahami2022blended, yang2022diffusion}, often through attention-based designs that encode spatial and temporal coherence. Although such methods improve perceptual consistency, they often fail to ensure accurate pixel-level correspondence, limiting their effectiveness in geometry-sensitive applications. An alternative direction leverages diffusion model distillation to directly supervise or optimize 3D representations, such as NeRFs or 3D Gaussian splats \cite{poole2023dreamfusion, metzer2022latentnerf, liu2024infusioninpainting3dgaussians}. While these approaches offer improved structural coherence, the generative nature of diffusion priors can lead to artifacts or unrealistic geometry in inpainted regions, highlighting the trade-off between photorealism and geometric fidelity in diffusion-based 3D scene editing.

\paragraph{Projection-based 3D Editing}
These methods employ depth-based projection to map information from a reference view (the anchor) to multiple target views in pixel or latent space, enabling manipulation of occluded or missing regions via inpainting, feature distillation, interpolation, or accumulation-based strategies \cite{wang2023zero123plusplus, liu2024infusioninpainting3dgaussians, yu2023dream3d, zhang2023multigaussian}. A core challenge lies in acquiring accurate and globally consistent depth maps for world-coordinate projections. Monocular depth estimators, while widely used \cite{ranftl2020dpt, liu2021adaptivebins}, often suffer from scale ambiguity, which limits their direct applicability for precise 3D projections. Previous work \cite{hu2024innout, lu2024view, wang2024gscream} addresses this by introducing a dedicated pretraining phase to obtain metric-consistent depth maps, allowing for more reliable anchor-to-target view warping.

\paragraph{Geometry-aware 3D Reconstruction}
Geometry-aware methods have become integral to 3D reconstruction tasks, particularly for few-shot reconstruction, novel view synthesis, and scene editing \cite{niemeyer2022regnerf, xu2023denseposediffusion, wang2023neuralangelo}. These approaches use geometric priors like depth maps, surface normals, or multi-view constraints to enforce spatial coherence and reduce hallucinations. However, monocular depth estimation suffers from scale ambiguity, complicating direct supervision. Most methods use L1 or smoothness-based regularization losses, which require aligning predicted depth maps to a common scale—a computationally expensive and error-prone step \cite{ranftl2020dpt, liu2021adaptivebins}. 


\section{Preliminaries}

\paragraph{3D Gaussian Splatting}
The 3D Gaussian representation serves as the foundational structure for modeling radiance fields. Each Gaussian is defined by a 3D position $\boldsymbol{\mu} \in \mathbb{R}^3$, an anisotropic scale matrix $\mathbf{S} \in \mathbb{R}^{3 \times 3}$, a rotation matrix $\mathbf{R} \in \mathbb{R}^{3 \times 3}$, a color vector $\mathbf{c} \in \mathbb{R}^3$, and an opacity value $\alpha \in [0, 1]$. The volumetric density is expressed using the covariance matrix $\boldsymbol{\Sigma} = \mathbf{R} \mathbf{S} \mathbf{S}^\top \mathbf{R}^\top$ as:

\begin{equation}
G(\mathbf{x}) = \exp\left(-\frac{1}{2}(\mathbf{x} - \boldsymbol{\mu})^\top \boldsymbol{\Sigma}^{-1} (\mathbf{x} - \boldsymbol{\mu})\right).
\end{equation}

Following the projection model in \cite{zwicker2001surface}, each 3D Gaussian is mapped onto the 2D image plane, producing an elliptical footprint. Final pixel colors are computed using volumetric compositing via alpha blending:

\begin{equation}
\hat{C} = \sum_{k=1}^K \mathbf{c}_k \alpha_k \prod_{j=1}^{k-1} (1 - \alpha_j),
\end{equation}

where $K$ is the number of Gaussians intersected along a camera ray. Initial positions are initialized using sparse SfM points \cite{Schonberger_2016_CVPR}. All Gaussian parameters are jointly optimized via photometric reconstruction loss, as described in \cite{kerbl3Dgaussians}.

\paragraph{Multiview Stereo Depth} Multiview stereo (MVS) methods aim to estimate depth maps that are consistent across multiple views by leveraging known camera parameters to geometrically project pixels from a reference image to several source views \cite{mvsnet}. These projections enable the construction of a cost volume, which is then processed to infer depth. While MVS approaches can achieve high accuracy, they are typically computationally expensive and struggle in textureless or occluded regions \cite{casmvsnet, patchmatchnet}. In contrast, deep stereo depth (DSD) estimation methods use neural networks to directly predict depth or disparity maps from image pairs \cite{mccnn, gcnet, ganet-cvpr2022, Wang_2024_CVPR, ye2025no}. Instead of relying on explicit geometric reasoning, DSD models learn to capture pixel-scene relationships through data-driven representations. Although this may sacrifice exact pixel-to-world alignment, DSD approaches maintain cross-view geometric consistency and offer much faster inference, making them ideal for real-time and large-scale 3D applications \cite{hitnet, deeppruner, acvnet}.

\section{Method}

\begin{figure}
    \centering
    \includegraphics[width=\linewidth]{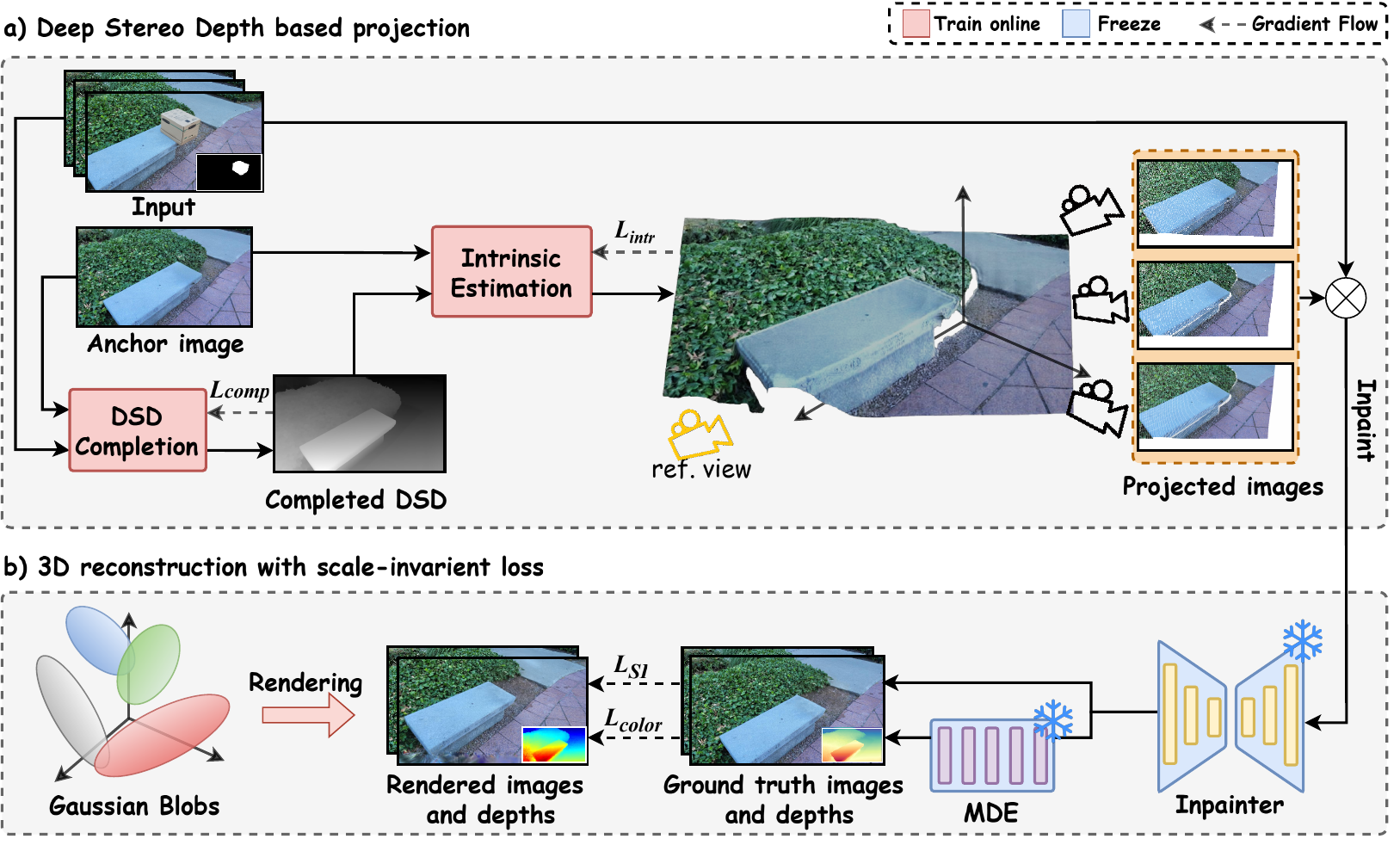}
    \caption{\textbf{Pipeline overview.} The first stage (a) performs stereo depth completion on the anchor view and estimates the implicit intrinsic matrix to enable accurate projection of inpainted content to other views. The second stage (b) uses a pretrained inpainting network to complete the masked regions in each projected view. The resulting multi-view images are then used for 3D reconstruction via Gaussian Splatting, guided by scale-invariant and photometric losses to ensure high-fidelity and geometrically consistent outputs.}
    \label{fig:pipeline}
\end{figure}

Given a set of multi-view posed images $\{I_i\}_{i=1}^N$ of a static real-world scene, along with camera poses derived from Structure-from-Motion (SfM), and binary masks $\{M_i\}_{i=1}^N$ identifying object regions, our goal is to produce a geometrically consistent and inpainted training set $\{\tilde{I}i\}_{i=1}^N$ for supervising 3D Gaussian Splatting (3DGS). These masks are assumed to be provided during training and can be obtained using established methods in prior works~\cite{spinnerf, lu2024view, yin2023ornerf}.

Given a selected anchor image $I_{0}$ and the input dataset, we first apply Deep Stereo Depth (DSD) Completion to obtain a completed deep stereo depth map, supervised by the completion loss $\mathcal{L}_{\text{comp}}$. This anchor view is then lifted into 3D using the completed DSD and an implicitly estimated intrinsic matrix $\tilde{K}$, optimized via the intrinsic loss $\mathcal{L}_{\text{intr}}$. The inpainted content from the anchor view is projected to the remaining views $\{I_i\}_{i=1}^{N-1}$, where it is combined with the original images and corresponding input masks to guide scene completion.

In the second stage, a pretrained inpainting model is then used to fill missing regions based on the geometry-aligned projections from the previous stage, resulting in a completed set of multi-view images $\{\tilde{I}i\}_{i=1}^N$. These completed views are used to supervise 3D reconstruction via Gaussian Splatting, guided by a scale-invariant depth loss $\mathcal{L}_{\text{SI}}$ and a photometric reconstruction loss $\mathcal{L}_{\text{color}}$, using monocular depth estimates from a pre-trained depth estimator. The full pipeline is illustrated in Fig.~\ref{fig:pipeline}.

The method section is organized into two main parts: Deep Stereo Depth-based Projection in Sec.~\ref{sec:dsd-projection}  and Scale-Invariant Depth Loss for Structure Guidance in Sec.~\ref{sec:scale-invariant-loss}. In the first part, we delve into the details of how Deep Stereo Depth Completion is applied and how to estimate the  intrinsic matrix to enable accurate depth-based projection. The second part focuses on the design and role of the scale-invariant depth loss.


\subsection{Deep Stereo Depth-Based Projection}
\label{sec:dsd-projection}
\paragraph{Stereo Depth Completion for Inpainted Regions} 
To efficiently lift the scene into 3D space for reliable projection, it is crucial to obtain dense depth maps that are consistent with multi-view geometry. To this end, we build upon prior work on Deep Stereo Depth (DSD), which rapidly provides dense depth estimations. However, since inpainted regions lack ground-truth geometry, an additional step is required to predict their depth values and ensure alignment with the reference view. Previous approaches \cite{lu2024view, wang2024gscream} have addressed similar challenges by applying global or region-specific linear transformations—typically involving a learned scale and offset—to monocular depth predictions. While such techniques can facilitate loss convergence, they often result in broken structure and misalignment, particularly near complex boundaries, due to their limited capacity to model spatially varying depth patterns. 

To overcome this limitation, we propose using a lightweight neural network that adaptively aligns the monocular depth predictions with the view-consistent estimated depth. This network refines the initial predicted mono-depth $d(\mathbf{p})$ by producing a corrected estimate $\mathcal{F}(d(\mathbf{p}))$. This design enables smoother geometric transitions and improves the accuracy of projections, especially in boundary regions. The alignment is supervised using a loss defined on the boundary of inpainted masks, which encourages local coherence between the predicted and estimated depths. Formally, the loss is given by:

\begin{equation}
\mathcal{L}_{\text{comp}} = \sum_{\mathbf{p} \in \partial \mathcal{M}} \left(\mathcal{F}(d(\mathbf{p})) - D(\mathbf{p})\right)^2,
\end{equation}

where $\partial \mathcal{M}$ denotes the set of pixels near the boundary of the inpainted region, $d(\mathbf{p})$ is the predicted mono-depth, $\mathcal{F}(d(\mathbf{p}))$ is the refined depth from the neural network, and $D(\mathbf{p})$ is the view-consistent depth at pixel $\mathbf{p}$.

\paragraph{Intrinsic Estimation for Projection} 
Deep Stereo Depth (DSD) methods typically produce implicit pixel-to-scene correspondences. To adapt DSD for projection-based tasks, we propose estimating an implicit camera intrinsic matrix $\tilde{K}$, which establishes an explicit mapping between views by aligning the predicted depth with a projective geometry framework. A natural supervision strategy involves projecting the anchor image to the target view using the predicted depth and $\tilde{K}$, followed by computing a photometric (RGB) reconstruction loss against the target image. However, this photometric loss provides only indirect supervision and lacks strong geometric constraints. Additionally, iterative projection operations are computationally expensive, which can lead to unstable training dynamics and inconsistent convergence.

To provide a more robust and geometrically grounded supervision signal for $\tilde{K}$, we alternatively leverage sparse 3D point clouds obtained from an initial pose estimation (e.g., via Structure-from-Motion). These 3D points are reprojected into the anchor view using its extrinsic parameters to generate sparse depth maps. Since point clouds are sparse, only a subset of pixels in the image have valid depth values derived from these 3D points. We denote this subset as $\mathcal{M}$, representing the set of pixels where sparse depth supervision is available. In parallel, the DSD-predicted depth is lifted to 3D using the estimated $\tilde{K}$, enabling a consistent geometric comparison. Specifically, the supervised loss to enforce consistency between the predicted and sparse depth is defined as:

\begin{equation}
\mathcal{L}_{\text{intr}} = \sum_{\mathbf{p} \in \mathcal{M}} \left( d(\mathbf{p}) - D_{sfm}(\mathbf{p}) \right)^2,
\end{equation}

where $d(\mathbf{p})$ denotes the depth predicted by the neural network at pixel $\mathbf{p}$, and $D_{sfm}(\mathbf{p})$ is the depth value obtained by projecting the corresponding 3D point into the anchor view using known extrinsics. After acquiring the implicit intrinsic matrix $\tilde{K}$, we leverage it to perform projection from the anchor image to other views using the predicted depth.

\subsection{Scale-Invariant Depth Loss for Structure Guidance}
\label{sec:scale-invariant-loss}
To bypass the need for explicit scale and offset regression \cite{wang2024gscream, chung2023depth}, we adopt a \textit{scale-invariant loss} that directly supervises monocular depth predictions based on relative scene geometry. Instead of enforcing absolute metric depth, the scale-invariant loss emphasizes structural consistency by operating in log-space, naturally handling global scale ambiguity. Following \cite{9710962}, the loss is formulated as:

\begin{equation}
\mathcal{L}_{\text{SI}} = \frac{1}{n} \sum_{i=1}^{n} \left( \log D_i - \log D_i^* \right)^2 - \frac{1}{n^2} \left( \sum_{i=1}^{n} \left( \log D_i - \log D_i^* \right) \right)^2,
\end{equation}

where $D_i$ and $D_i^*$ denote the rendered and predicted depth at pixel $i$, and $n$ is the number of valid pixels. This formulation ensures that uniform depth scaling does not affect the supervision signal, encouraging the model to adapt local depth variations crucial for maintaining multi-view consistency and preserving 3D structure. Compared to direct L1 supervision, scale-invariant losses relax the constraint on absolute depth values, promoting geometric consistency while enhancing robustness across varying data conditions. This enables the supervision signal to guide the entire reconstruction process beyond the initial structure formation, maintaining structural coherence throughout the reconstruction.

\paragraph{Training Objective}
The overall training loss jointly supervises both geometric and photometric consistency by combining a scale-invariant depth loss with a color reconstruction loss. The color term follows the formulation from the original 3D Gaussian Splatting method~\cite{kerbl3Dgaussians}, using a weighted combination of $\mathcal{L}_1$ and SSIM losses. The complete objective is defined as:

\begin{equation}
\mathcal{L}_{\text{total}} = \lambda_{\text{depth}} \mathcal{L}_{\text{SI}} +  \lambda_{\text{color}} \mathcal{L}_{1} + \lambda_{\text{ssim}} \mathcal{L}_{\text{SSIM}},
\end{equation}

where $\lambda_{\text{depth}}$, $\lambda_{\text{color}}$, and $\lambda_{\text{ssim}}$ are scalar weights that balance the contributions of the depth, color, and structural similarity terms in the overall loss.

\section{Experiment}
\subsection{Evaluation Setting}
\label{exp:setting}
\paragraph{Dataset:}
To assess the performance of our approach, we utilize the SPIn-NeRF \cite{spinnerf} dataset, a well-established benchmark commonly used in 3D scene editing research. This dataset features ten diverse real-world scenes, including three indoor and seven outdoor environments. For each scene, there are 60 training images that include unwanted objects and 40 testing images where these objects have been removed. All 100 images are accompanied by accurate camera pose data, and each frame consists of a binary mask highlighting the unwanted object’s location. In line with previous methods \cite{hu2024innout, wang2024gscream, lu2024view, cao2024mvinpainter}, we standardize all image resolutions to 1008 × 567 for fair comparison.
\paragraph{Baseline:}
In our work, we compare our method against current state-of-the-art approaches that address the same challenge of object removal and inpainting in 3D scenes, covering both NeRF-based and Gaussian Splatting-based techniques. Specifically, we evaluate against SPIn-NeRF \cite{spinnerf}, GScream \cite{wang2024gscream}, In-and-Out \cite{hu2024innout}, MVInpainter \cite{cao2024mvinpainter} and LaMask \cite{spinnerf}—the latter serving as a baseline to assess the performance of using only LaMa \cite{suvorov2022resolution} for inpainting. To ensure fairness, we utilize the original source code and configurations released by the authors of each method and run all experiments on a single NVIDIA RTX 3090 GPU with 24 GB of memory.
\paragraph{Metrics:}
For quantitative evaluation, we follow prior works by employing PSNR \cite{gonzalez2002digital} and LPIPS \cite{zhang2018unreasonable} to measure the reconstruction quality of individual views. These metrics provide insight into both pixel-level accuracy and perceptual similarity between the rendered outputs and the ground truth images. In addition to view-wise reconstruction quality, we assess multi-view consistency using LoFTR \cite{sun2021loftr}, a robust feature matching technique that helps quantify the structural coherence across different viewpoints. All metrics are computed within the object mask regions provided in the test set, as these areas are most affected by the object removal and inpainting processes.
\subsection{View Projection Results}
\begin{figure}
    \centering
    \renewcommand{\arraystretch}{0.95} 

    \begin{tabular}{@{}c@{\hspace{6pt}}c@{}c@{\hspace{6pt}}c@{}c@{}}
        \includegraphics[width=0.19\textwidth]{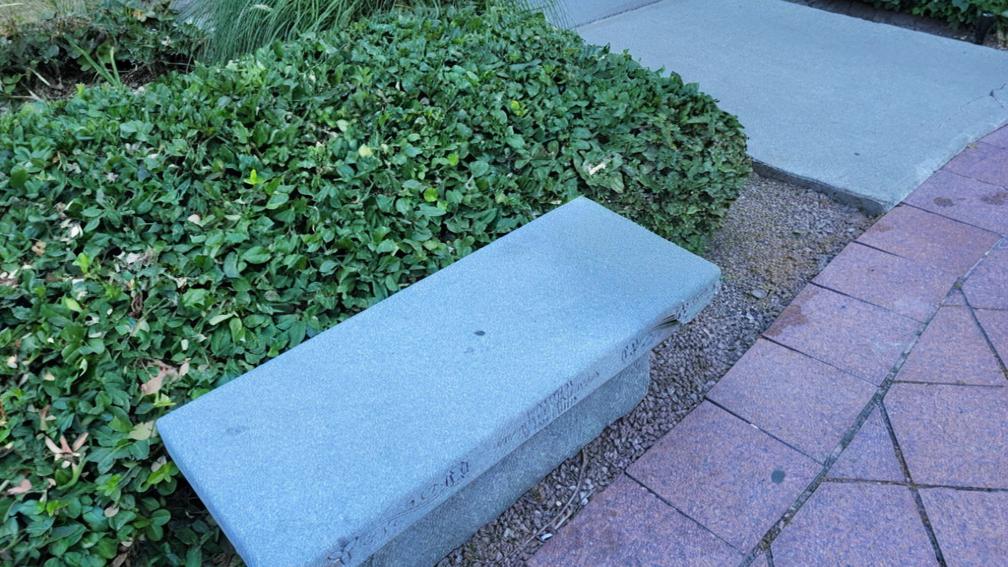} &
        \includegraphics[width=0.19\textwidth]{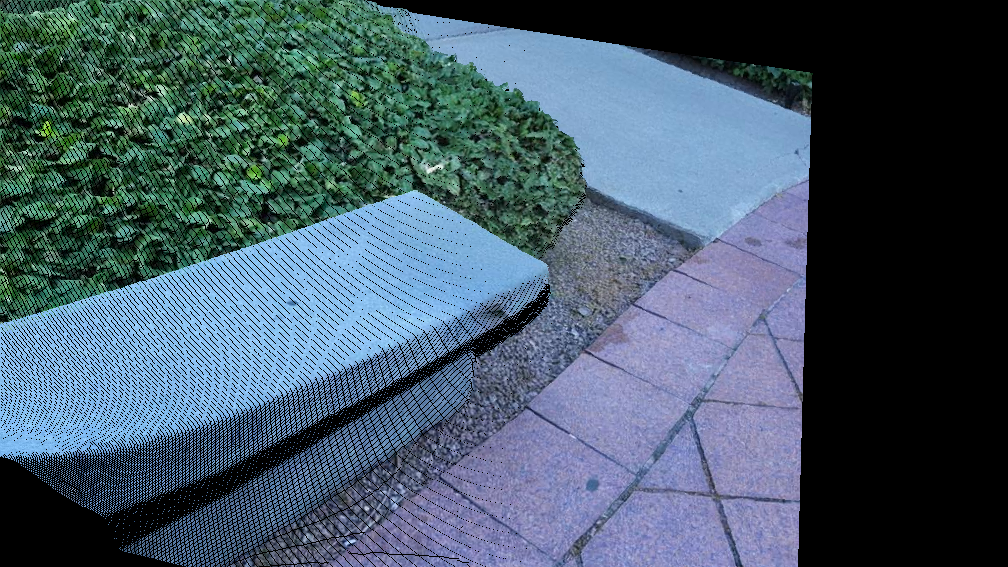} &
        \includegraphics[width=0.19\textwidth]{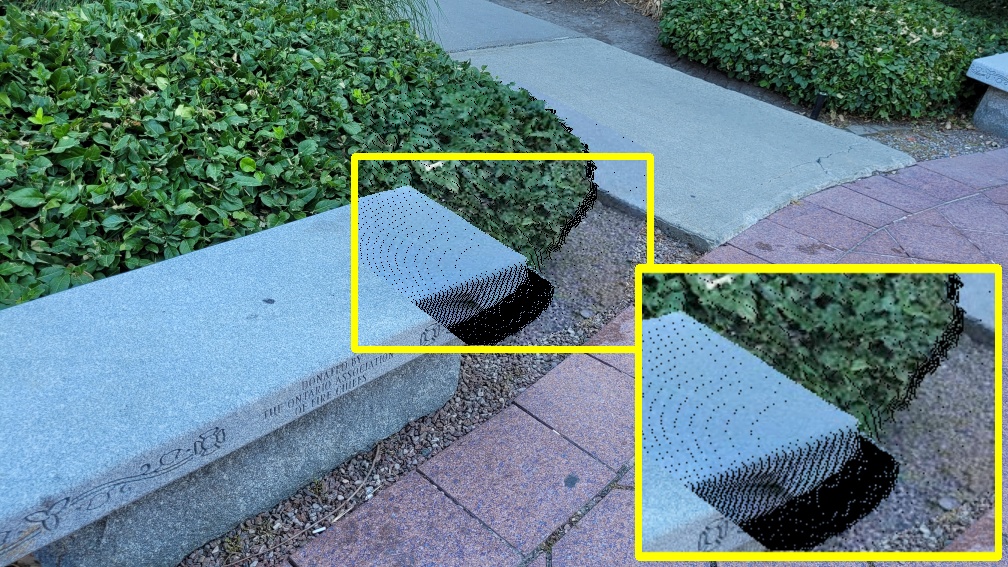} &
        \includegraphics[width=0.19\textwidth]{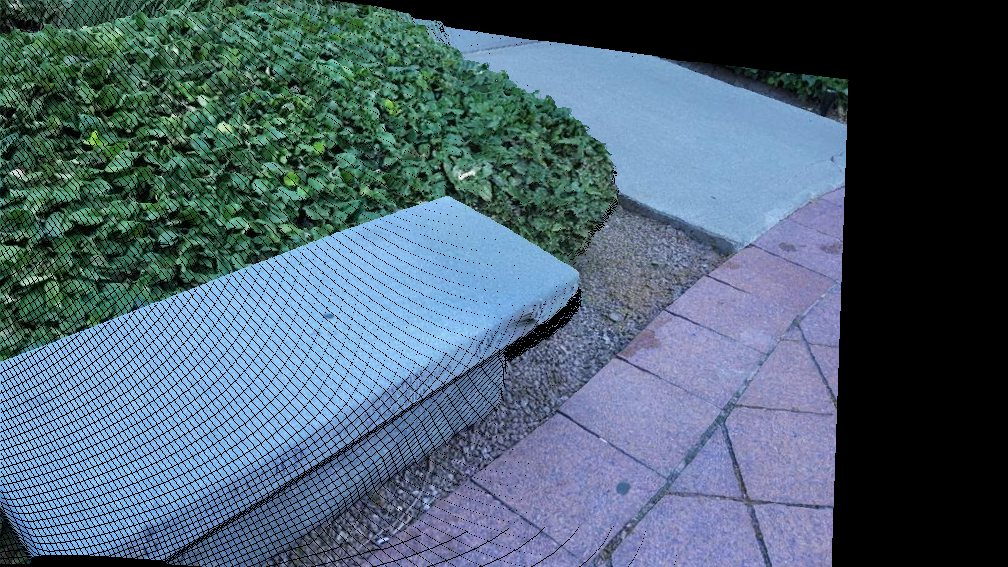} &
        \includegraphics[width=0.19\textwidth]{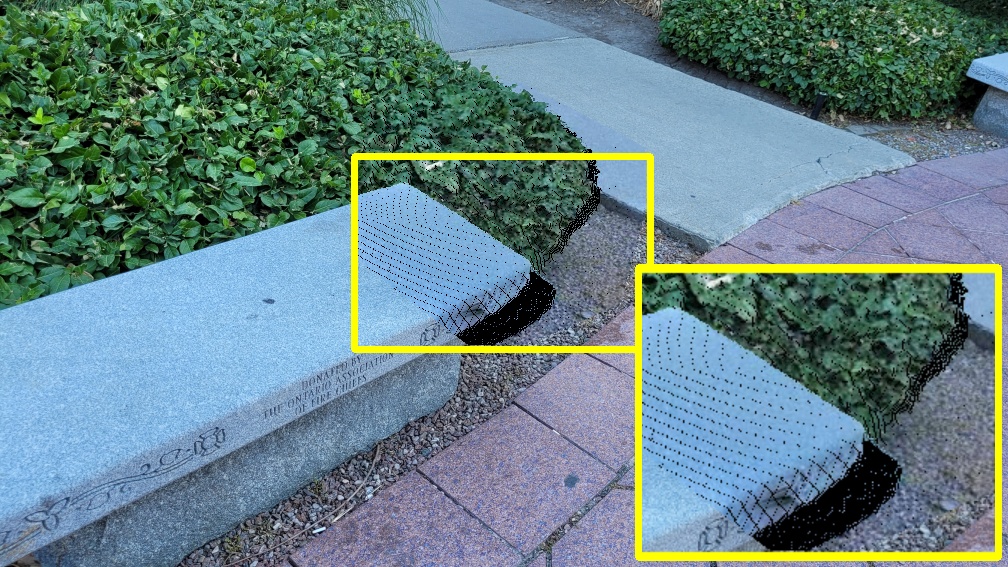} \\[-2pt]

        \includegraphics[width=0.19\textwidth]{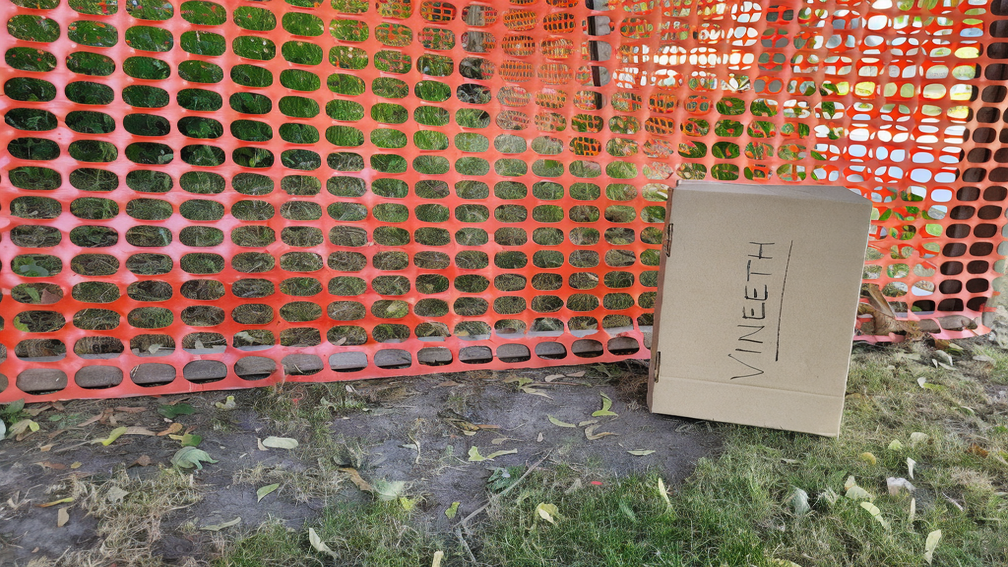} &
        \includegraphics[width=0.19\textwidth]{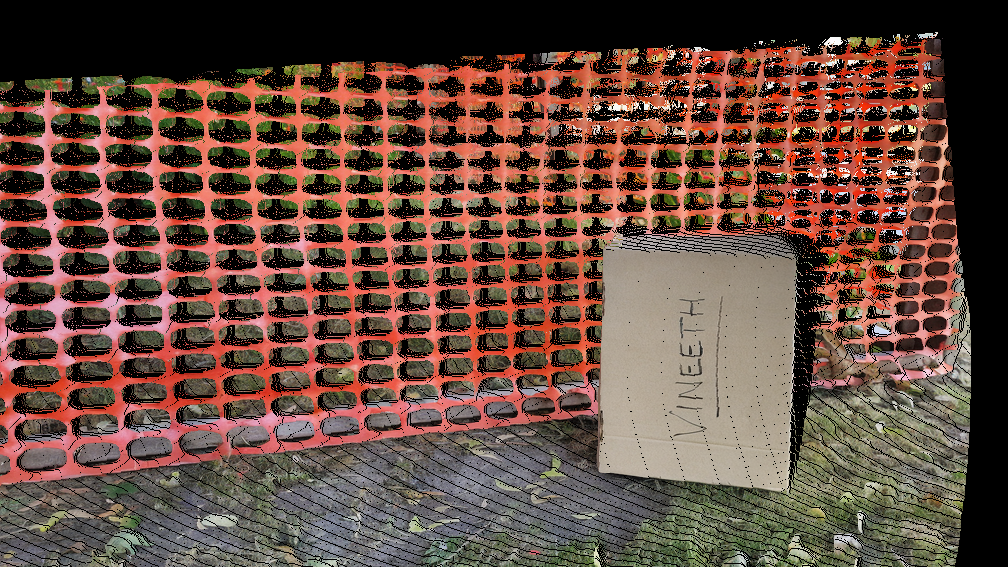} &
        \includegraphics[width=0.19\textwidth]{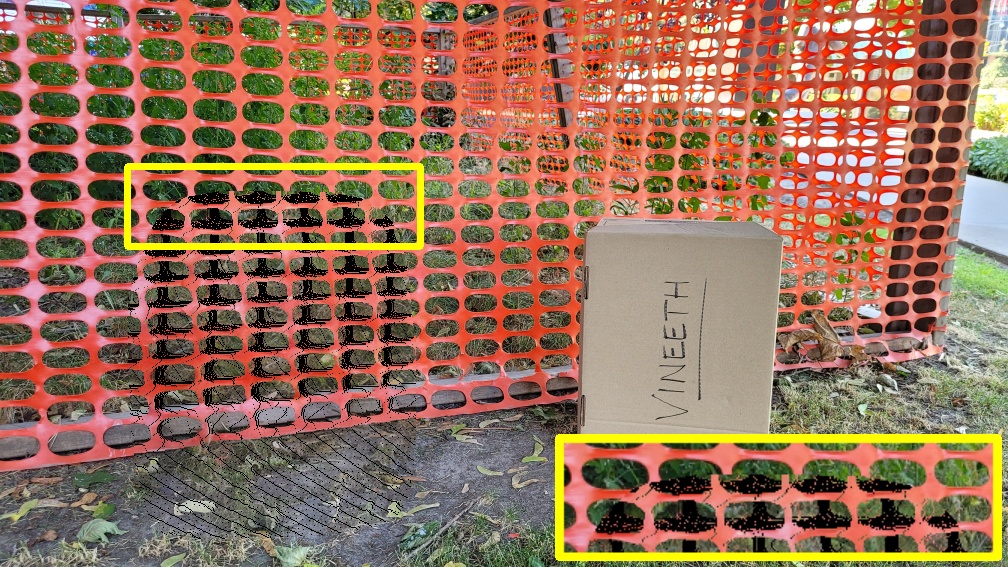} &
        \includegraphics[width=0.19\textwidth]{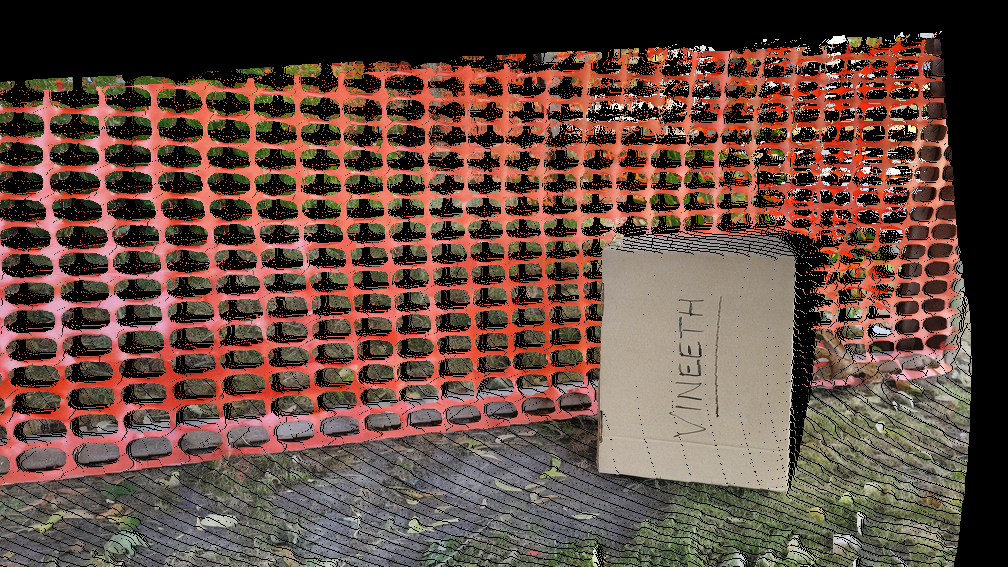} &
        \includegraphics[width=0.19\textwidth]{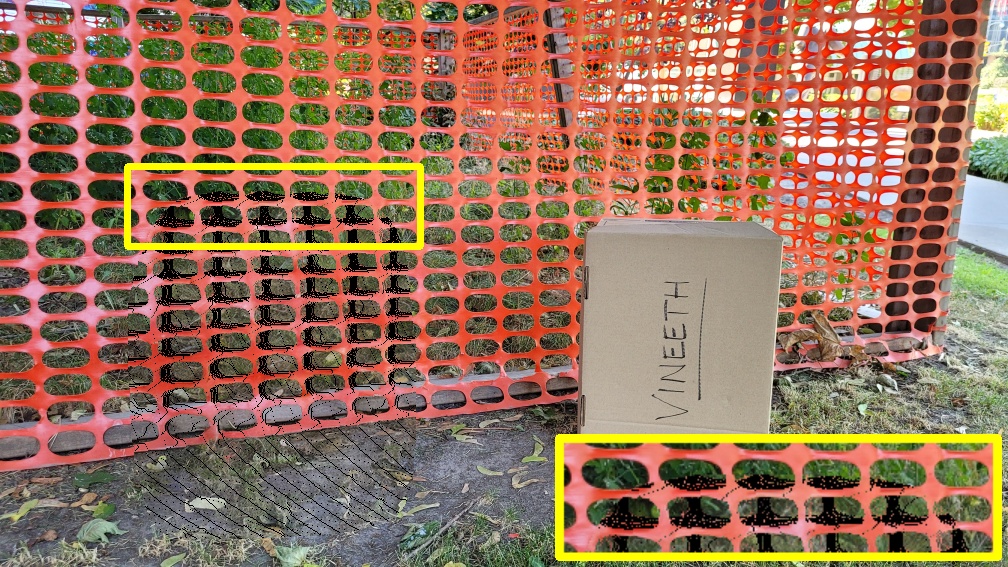} \\[-2pt]

        \includegraphics[width=0.19\textwidth]{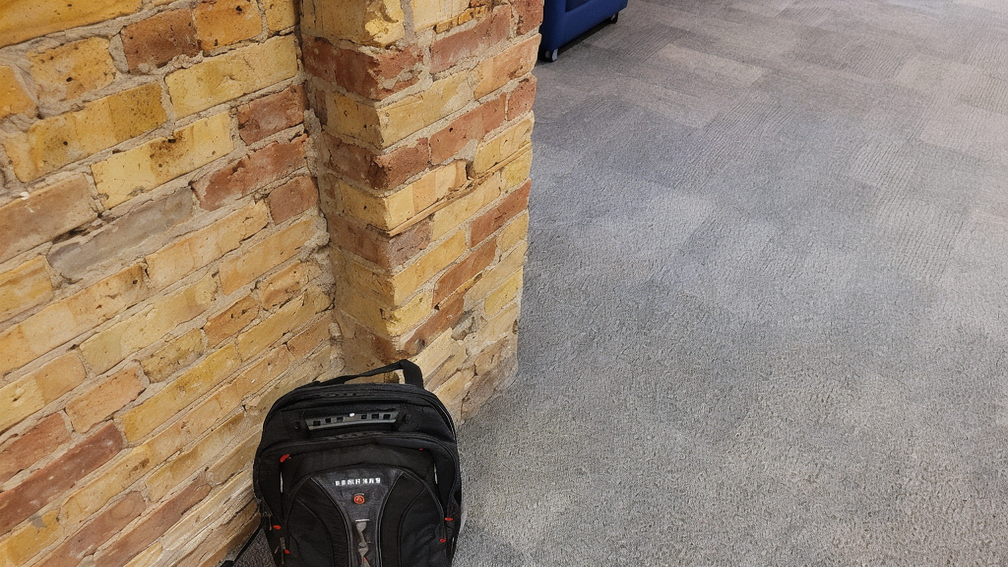} &
        \includegraphics[width=0.19\textwidth]{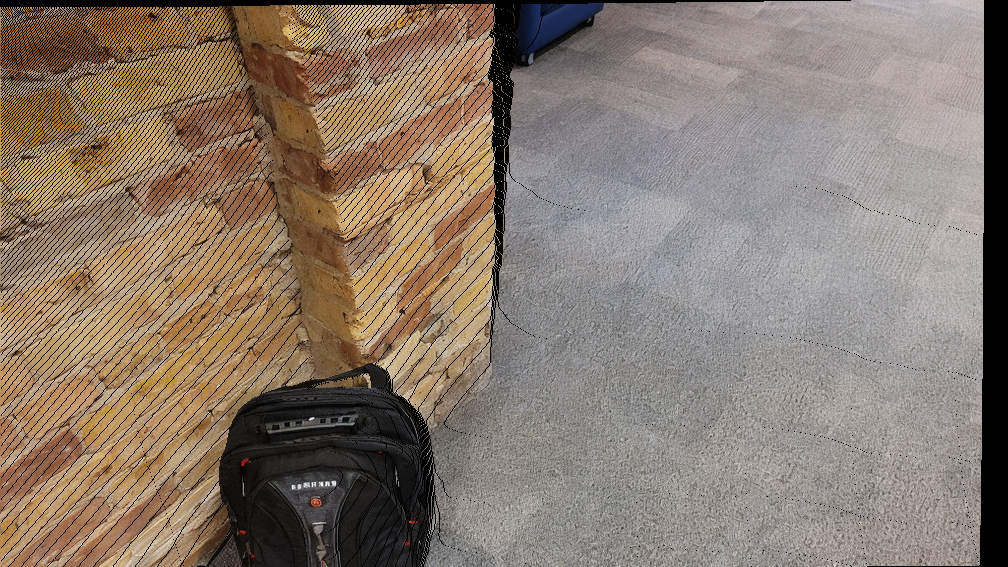} &
        \includegraphics[width=0.19\textwidth]{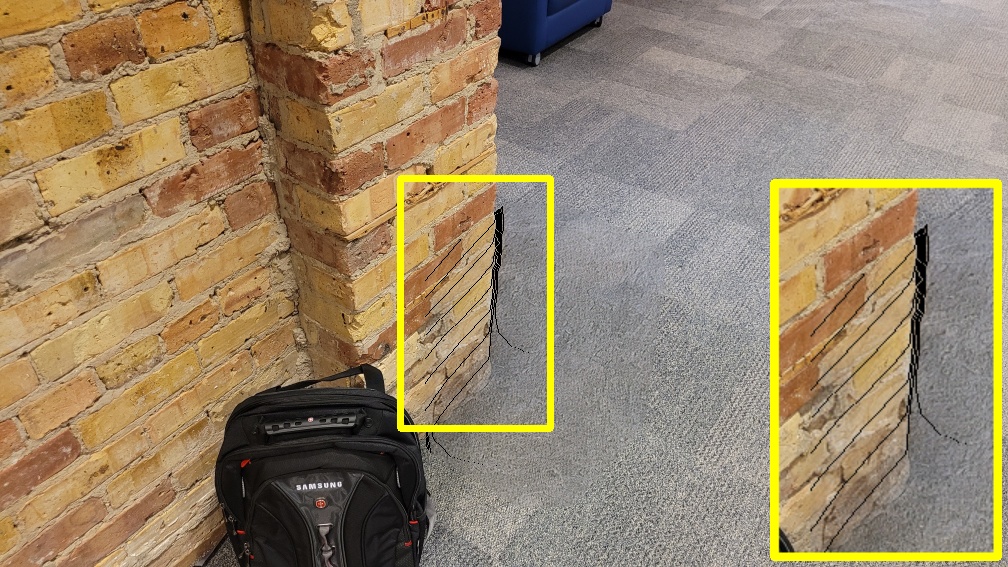} &
        \includegraphics[width=0.19\textwidth]{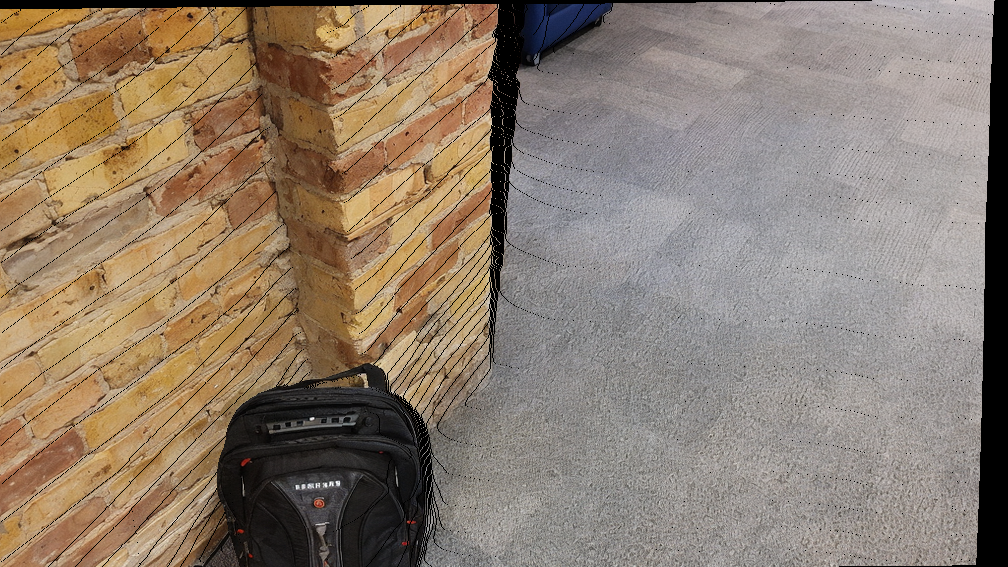} &
        \includegraphics[width=0.19\textwidth]{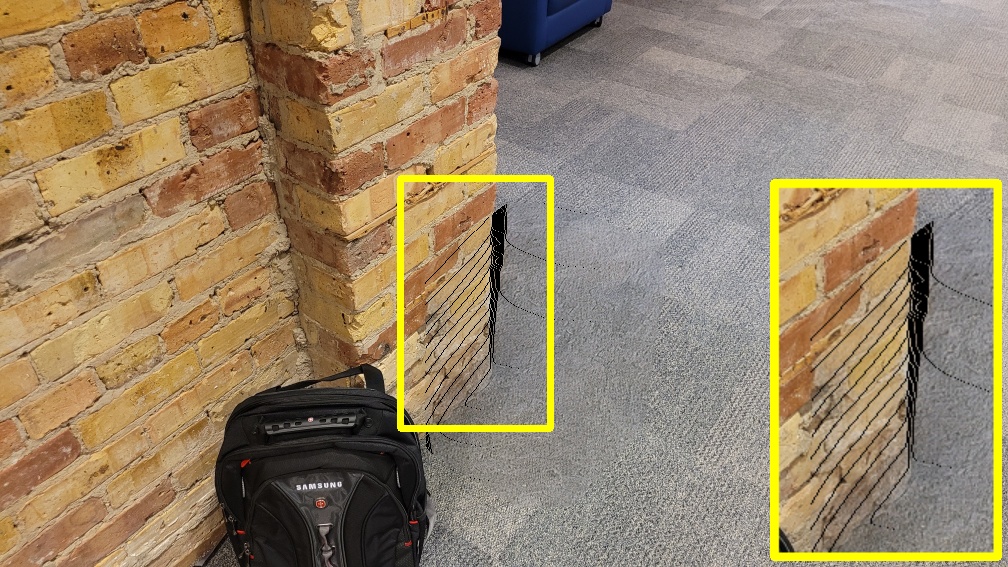} \\

        \multicolumn{1}{c}{(a) Anchor image} &
        \multicolumn{2}{c}{(b) Scaled-monodepth} &
        \multicolumn{2}{c}{(c) Ours}
    \end{tabular}

    \caption{Qualitative comparison of view projections from the (a) reference image using (b) scaled monocular depth and (c) our proposed method. Each result shows both unmasked (left) and masked (right) projections.}
    \label{fig:grid_subfig}
\end{figure}

Results of the projection stage are qualitatively compared in Fig. \ref{fig:grid_subfig}. We considered our method with representative one using scaled-monodepth. The first column (a) presents the anchor view used for projecting to the target view. The second (b) and third (c) parts illustrate the corresponding projection results using scaled monocular depth estimation: (b) shows the results from previous methods \cite{wang2024gscream, lu2024view, hu2024innout}, while (c) presents the results from our proposed method. These are shown with and without applying the original scene content to the surrounding masked regions. Notably, the presence of black regions in some projections highlights areas where information in the anchor view is insufficient to support accurate reconstruction in the target view. In complex scenes, projections based on scaled monocular depth often suffer from misalignments, especially around object boundaries, and exhibit fragmented geometric structures. In contrast, our approach produces more coherent and robust projections, seamlessly integrating object-removed regions with their surrounding context while preserving geometric integrity. This results in improved consistency and structural cohesion across all evaluated scenes. 

\subsection{Main Results}
\begin{table}
\centering
\caption{Quantitative comparison of different methods on masked regions.}
\begin{tabular}{lccccc}
\hline
\textbf{Method} & \textbf{PSNR $\uparrow$} & \textbf{LPIPS $\downarrow$} & \textbf{LoFTR $\uparrow$} & \textbf{Training Time (hrs) $\downarrow$} \\
\hline
LaMask & 16.301 & 0.680 & 95.108 & \textasciitilde 3.00 \\
SPIn-NeRF & 16.027 & 0.558 & 155.044 & \textasciitilde 3.00 \\
MV-Inpainter & 16.320 & 0.638 & 87.262 & -- \\
In-and-Out & 13.197 & 0.628 & 53.312 & \textasciitilde 1.70 \\
GScream & 15.749 & 0.531 & 215.166 & \textasciitilde 1.20 \\
Ours & \textbf{16.385} & \textbf{0.524} & \textbf{222.428} & \textbf{\textasciitilde 0.42} \\
\hline
\end{tabular}
\label{tab:quantitative}
\end{table}

\begin{figure}
\centering
\setlength{\tabcolsep}{1pt}
\renewcommand{\arraystretch}{1.0}

\begin{tabular}{ccccccc}
{Original} & {SPIn-NeRF} & {MV-Inpainter} & {In-and-Out} & {GScream} & \textbf{Ours} & \\
\includegraphics[width=0.15\linewidth]{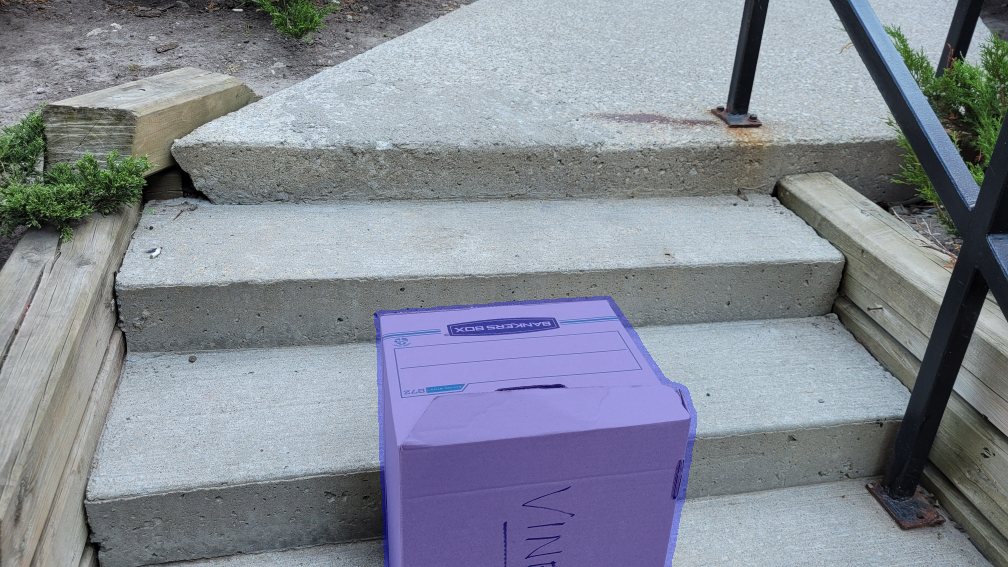} &
\includegraphics[width=0.15\linewidth]{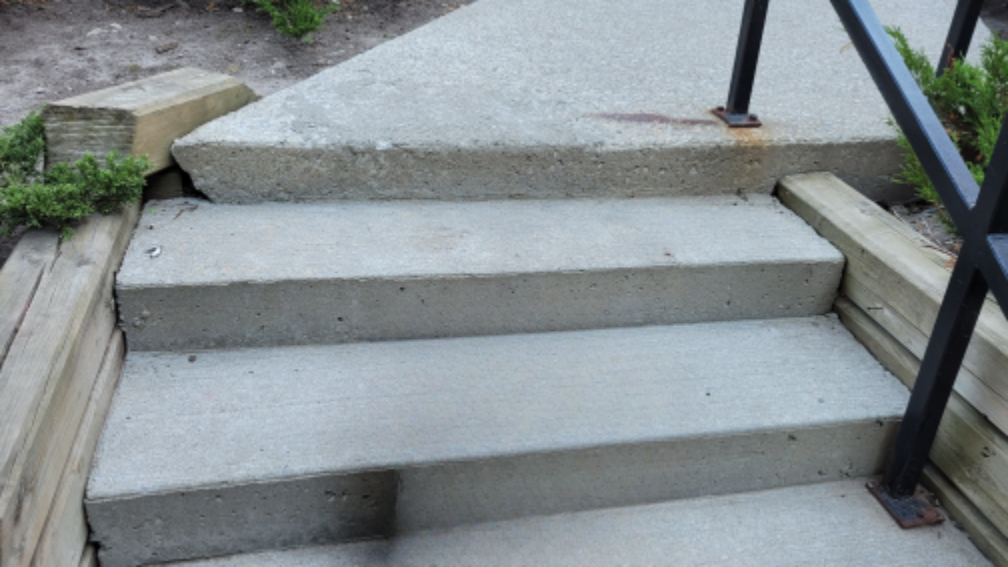} &
\includegraphics[width=0.15\linewidth]{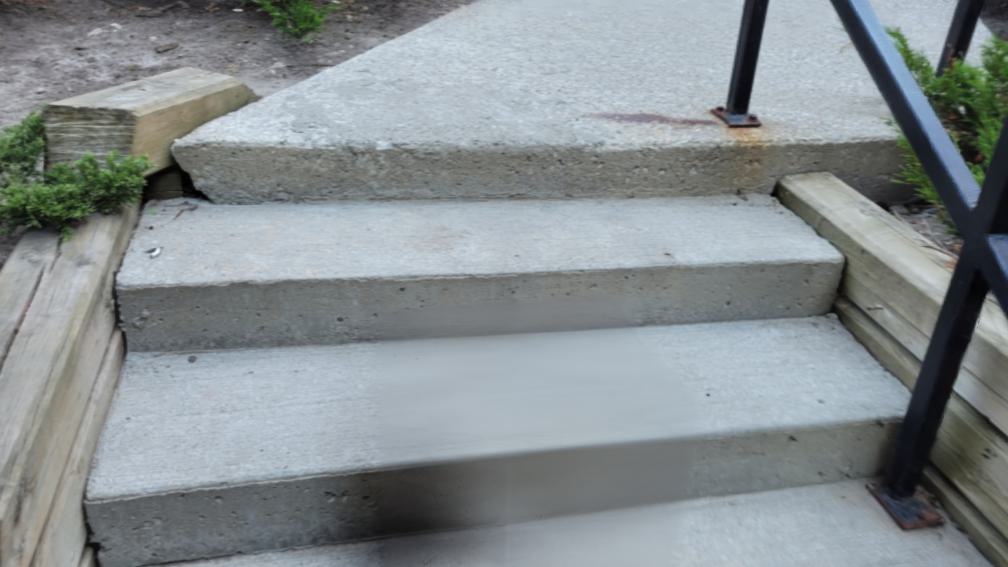} &
\includegraphics[width=0.15\linewidth]{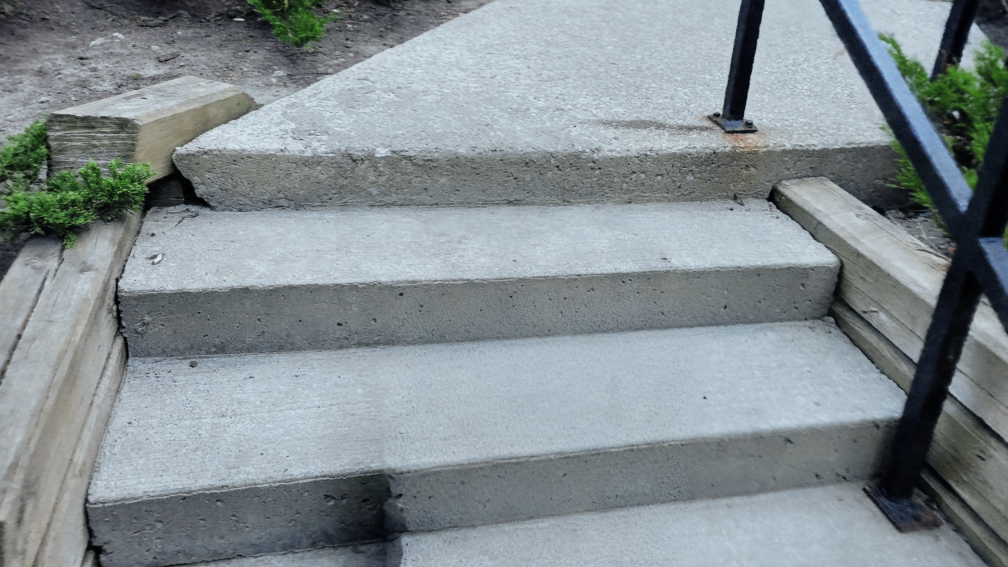} &
\includegraphics[width=0.15\linewidth]{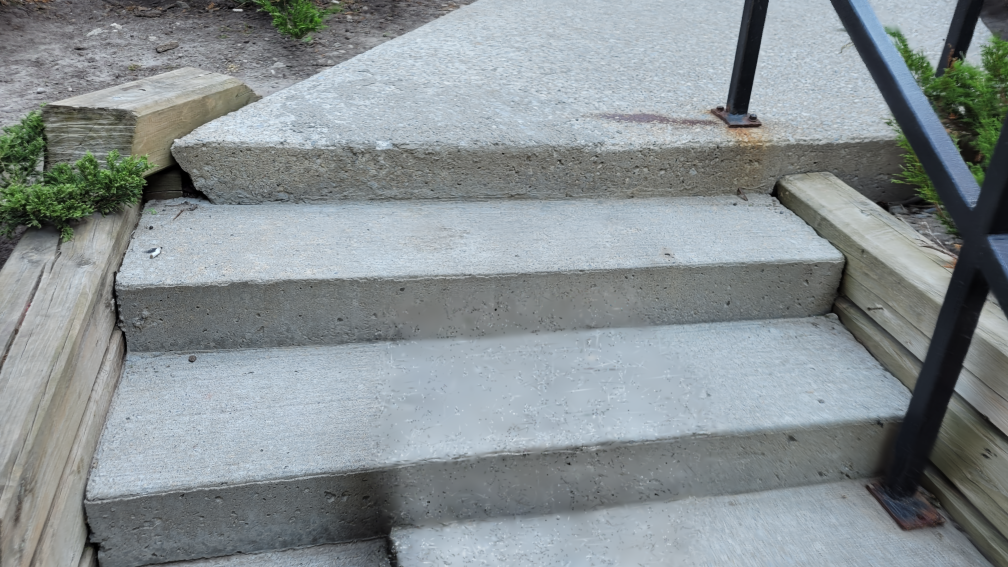} &
\includegraphics[width=0.15\linewidth]{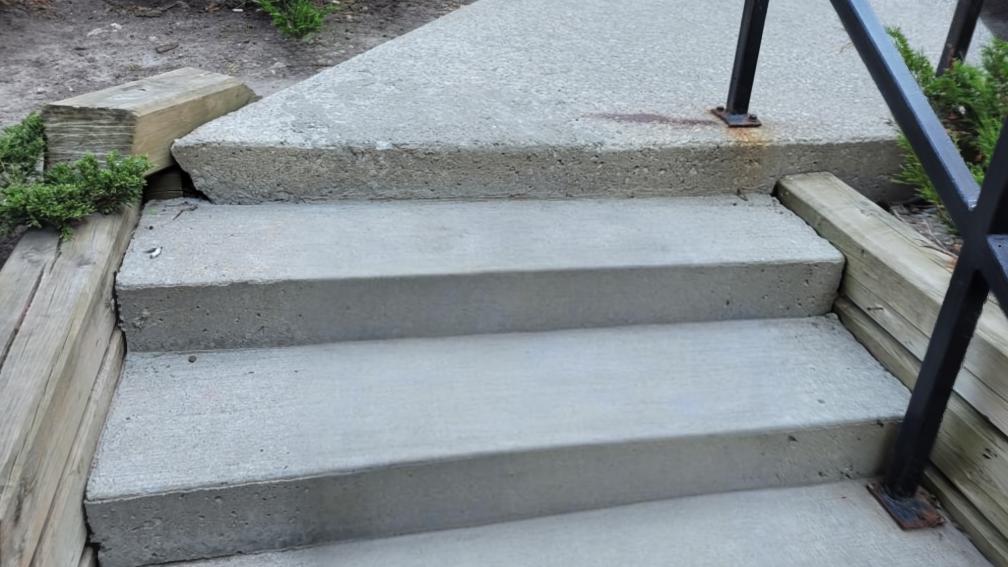} &
\rotatebox{90}{{View 1}} \\
\parbox[t]{0.15\linewidth}{\vspace{-7.0ex}\centering{(a) Scene 1}} &
\includegraphics[width=0.15\linewidth]{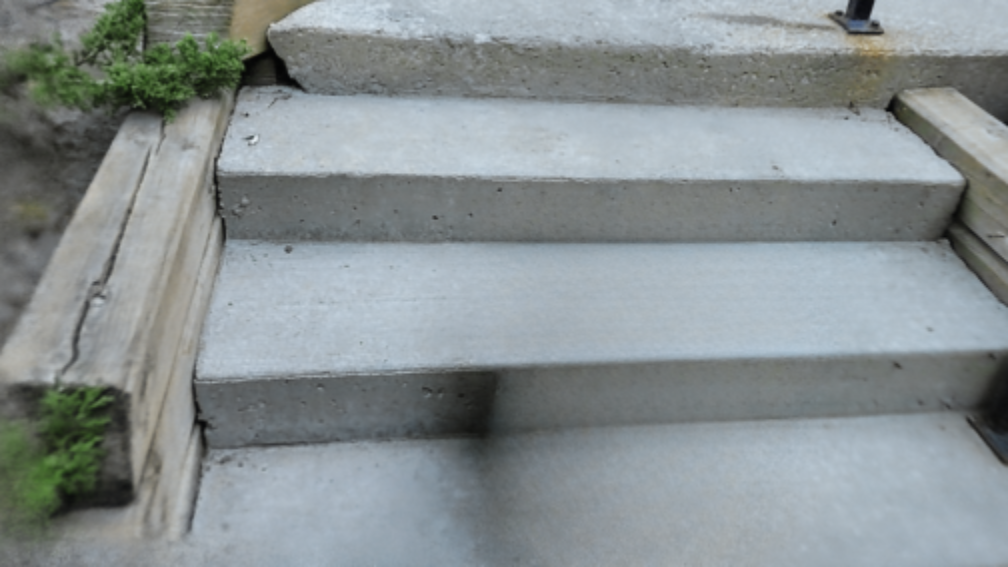} &
\includegraphics[width=0.15\linewidth]{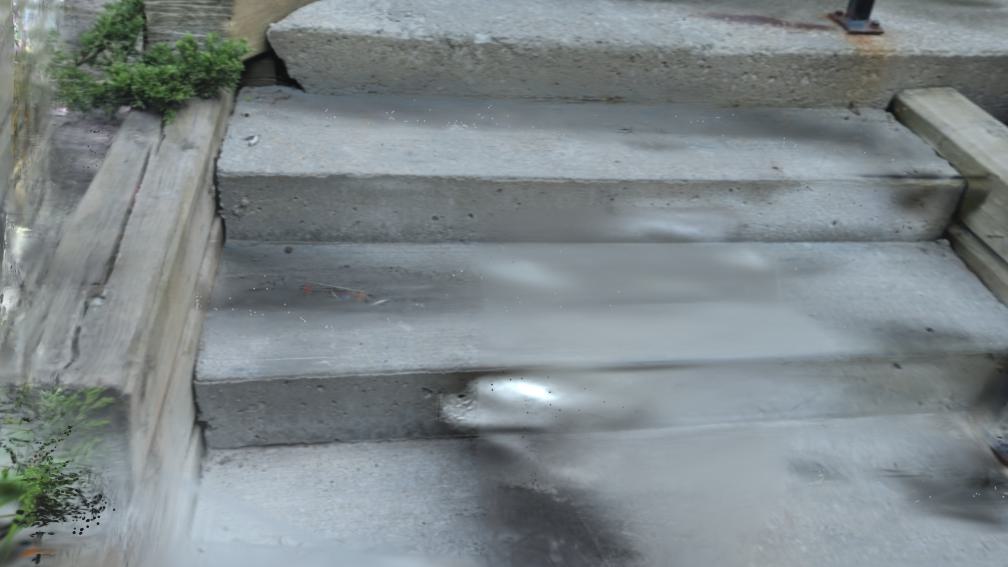} &
\includegraphics[width=0.15\linewidth]{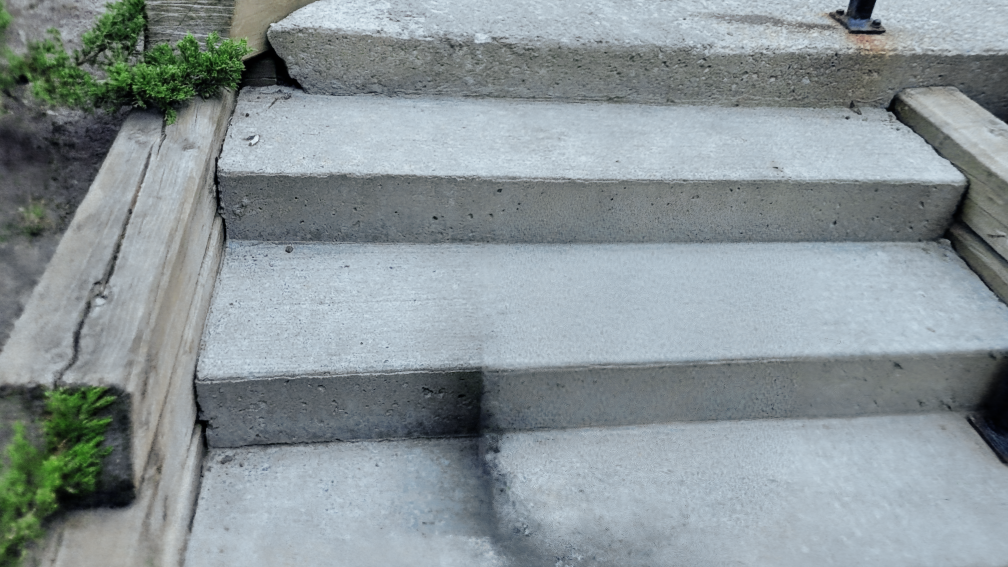} &
\includegraphics[width=0.15\linewidth]{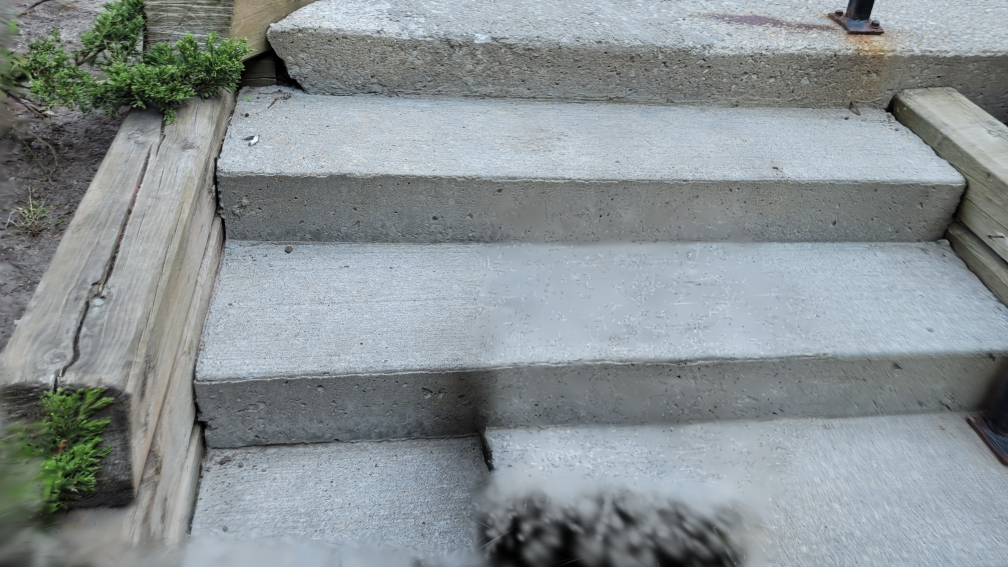} &
\includegraphics[width=0.15\linewidth]{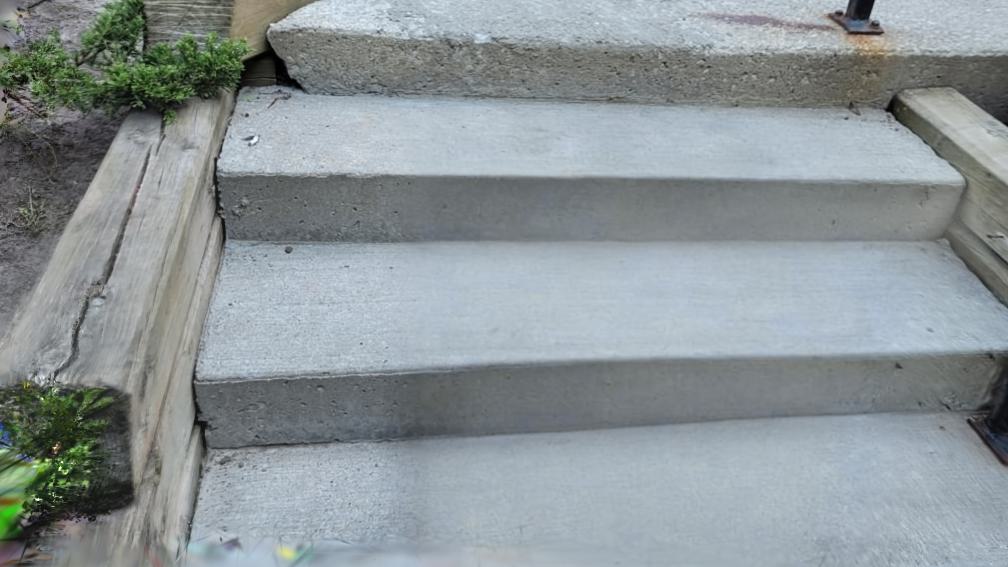} &
\rotatebox{90}{{View 2}} \\

\includegraphics[width=0.15\linewidth]{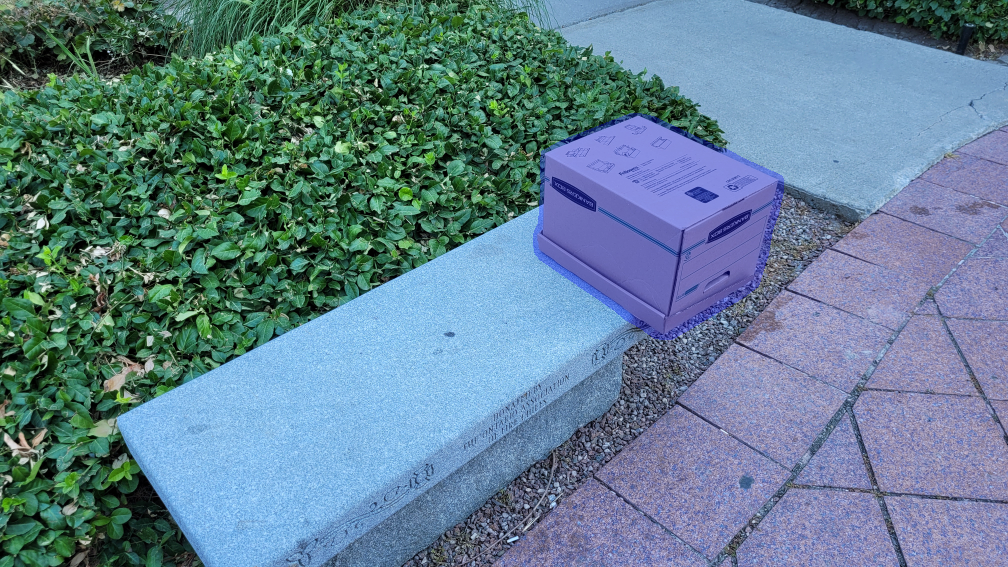} &
\includegraphics[width=0.15\linewidth]{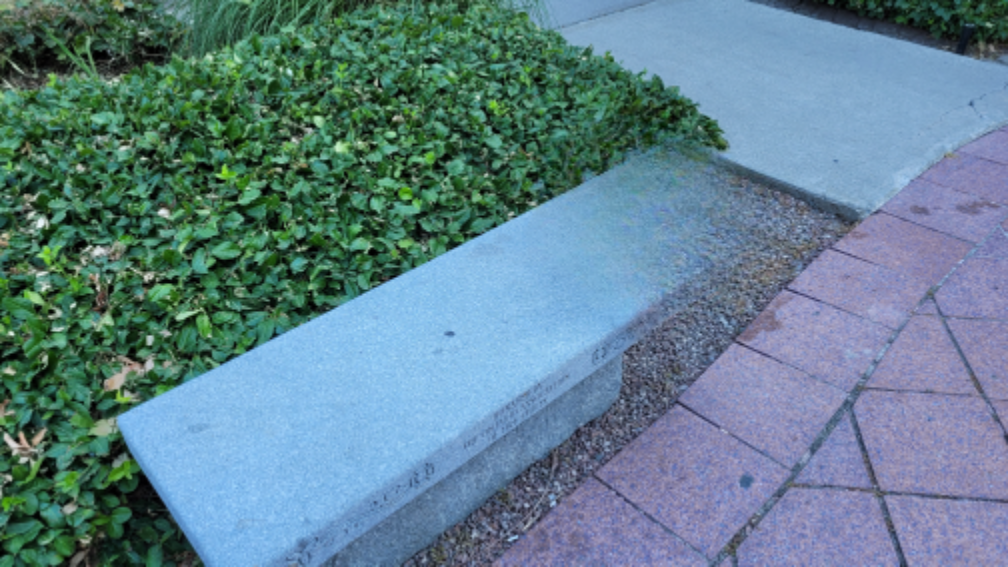} &
\includegraphics[width=0.15\linewidth]{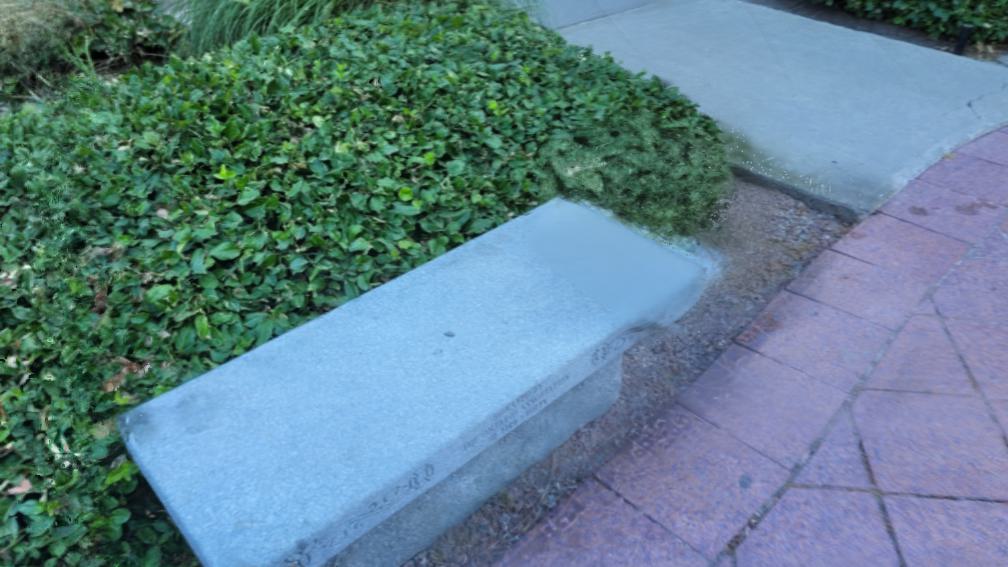} &
\includegraphics[width=0.15\linewidth]{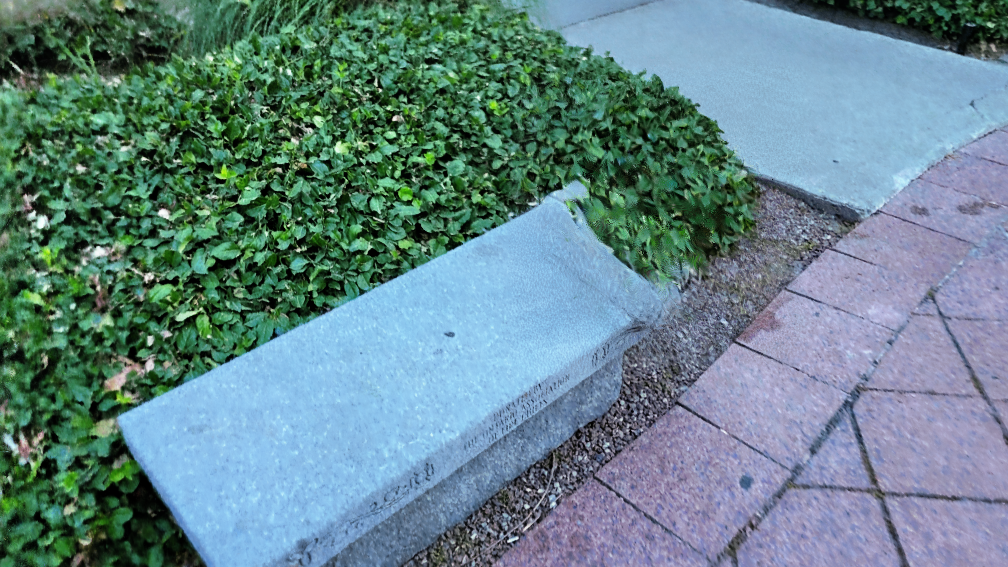} &
\includegraphics[width=0.15\linewidth]{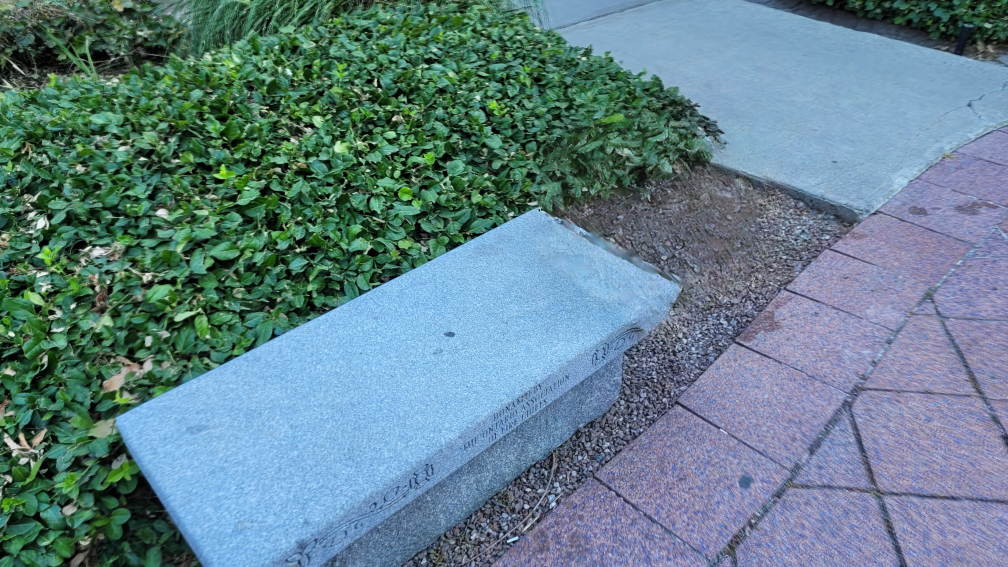} &
\includegraphics[width=0.15\linewidth]{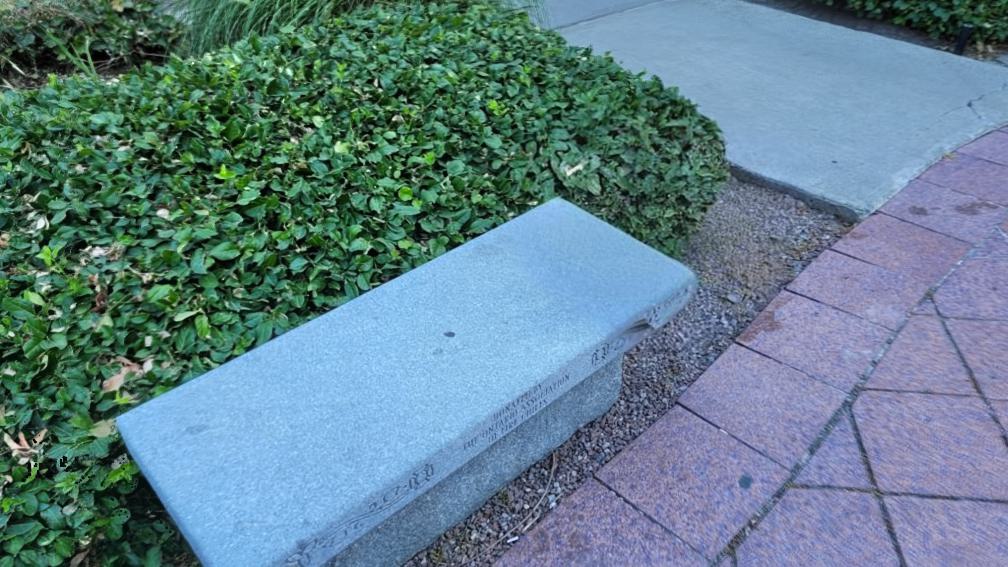} &
\rotatebox{90}{{View 1}} \\
 \parbox[t]{0.15\linewidth}{\vspace{-7.0ex}\centering{(b) Scene 2}} &
\includegraphics[width=0.15\linewidth]{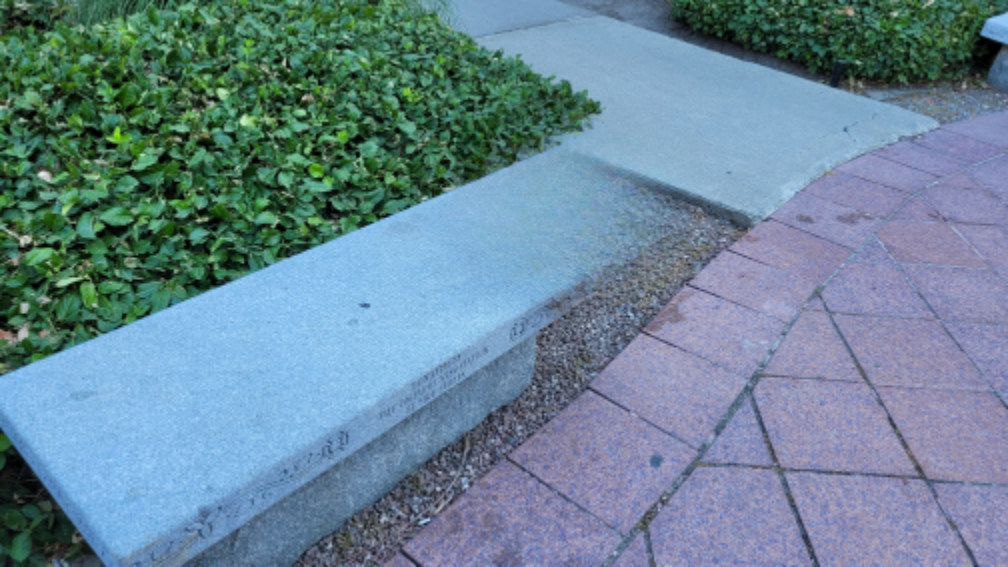} &
\includegraphics[width=0.15\linewidth]{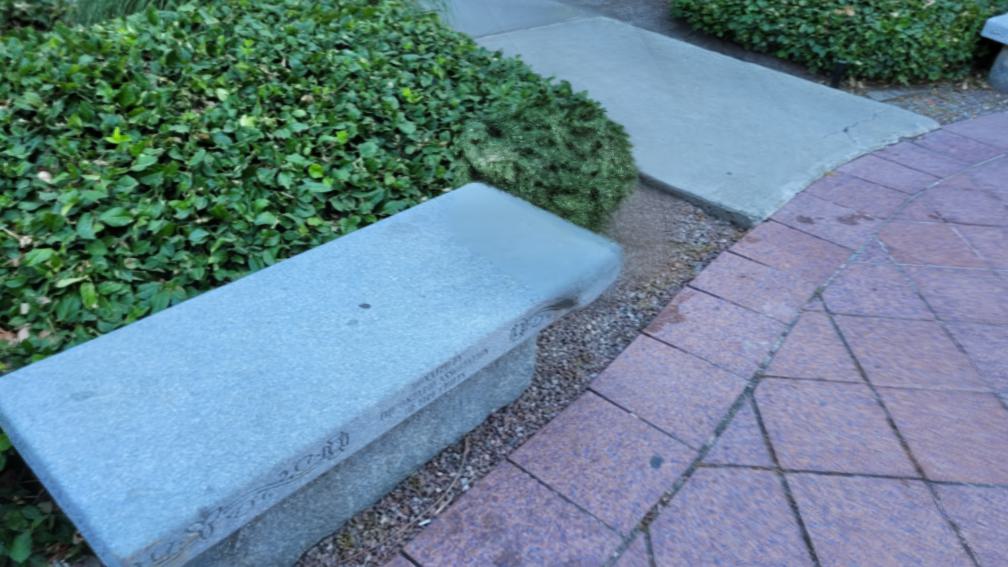} &
\includegraphics[width=0.15\linewidth]{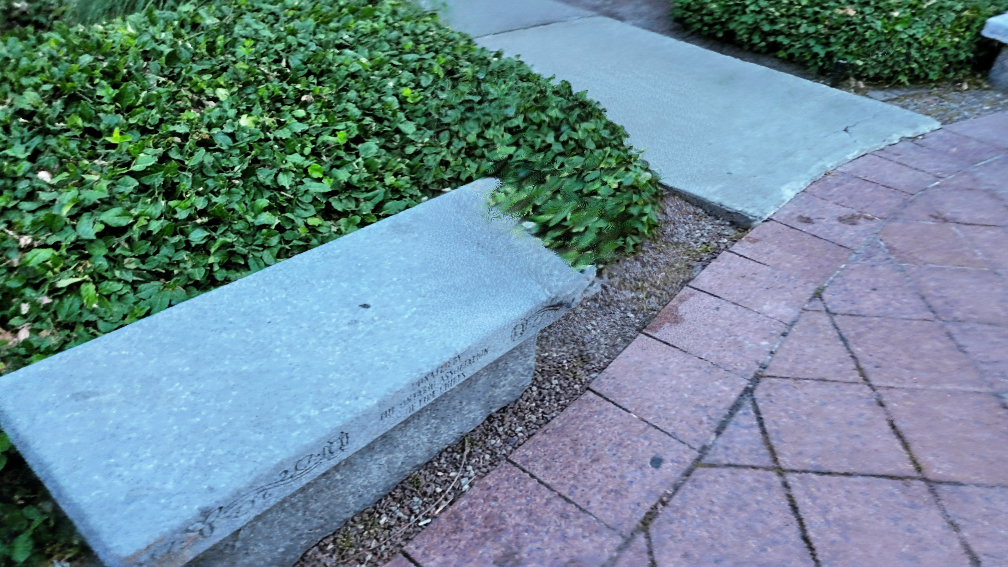} &
\includegraphics[width=0.15\linewidth]{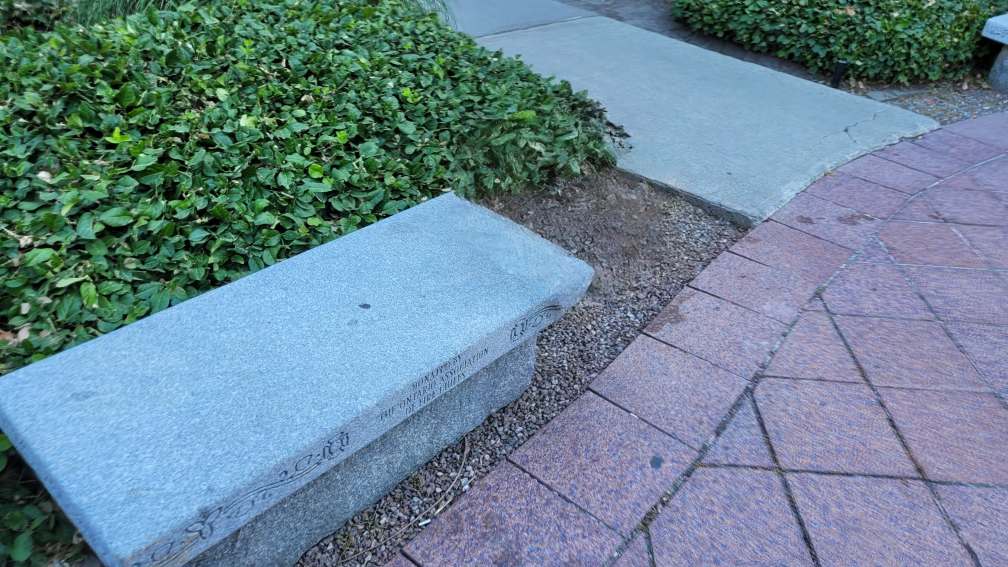} &
\includegraphics[width=0.15\linewidth]{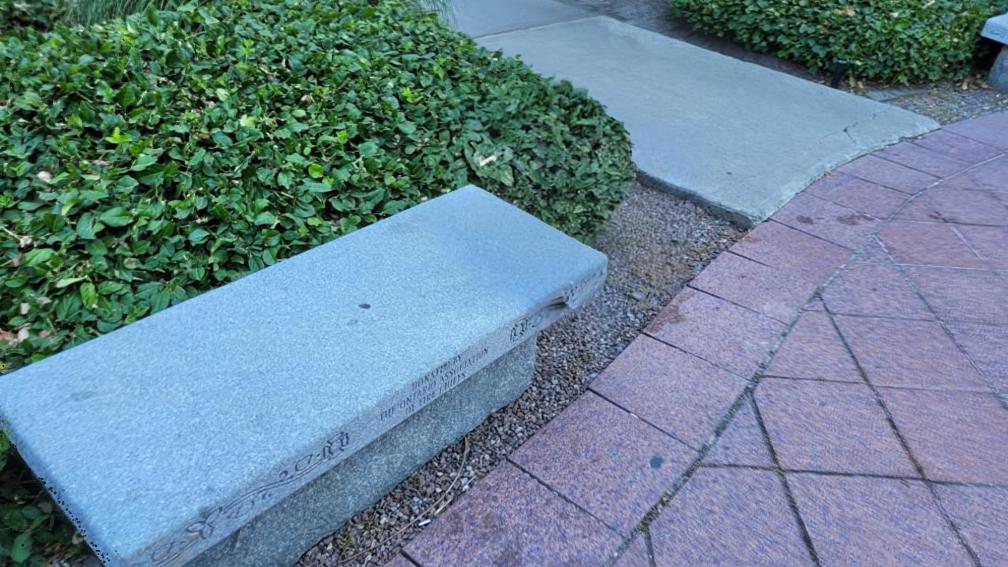} &
\rotatebox{90}{{View 2}} \\

\includegraphics[width=0.15\linewidth]{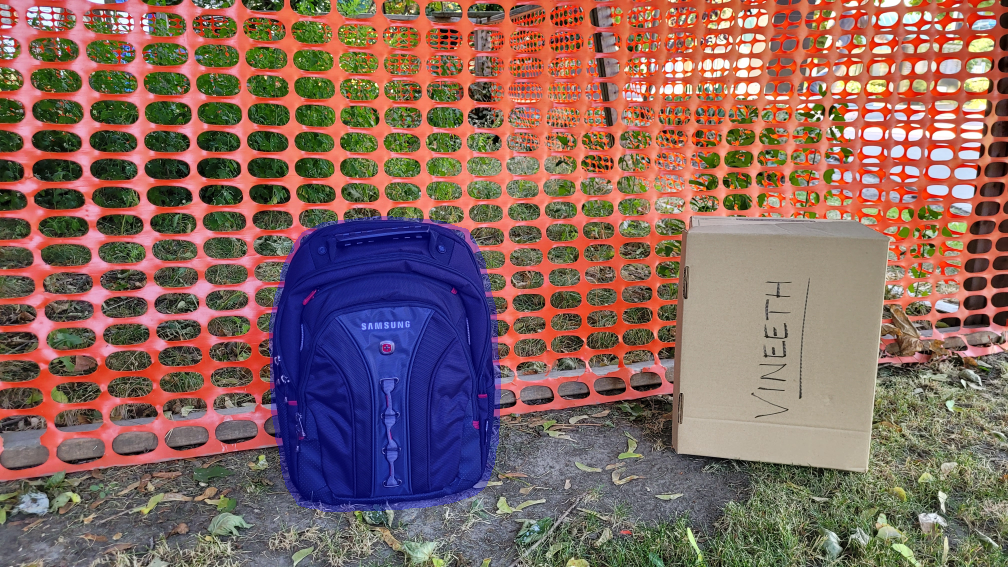} &
\includegraphics[width=0.15\linewidth]{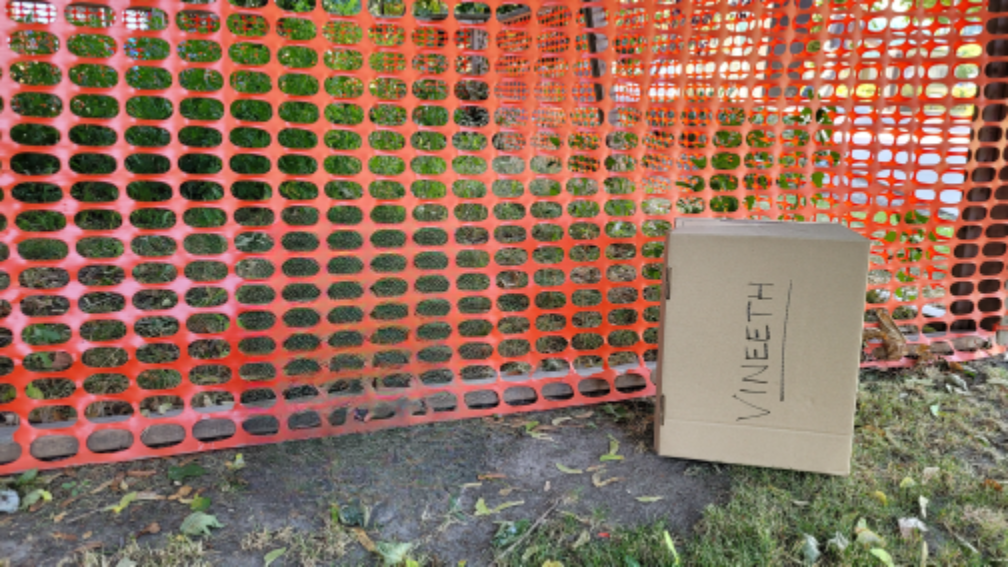} &
\includegraphics[width=0.15\linewidth]{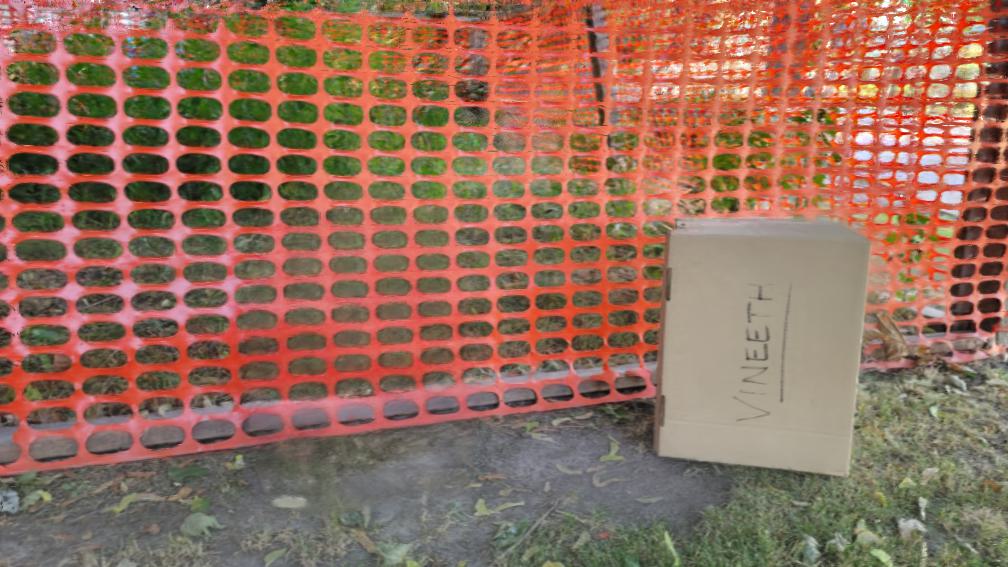} &
\includegraphics[width=0.15\linewidth]{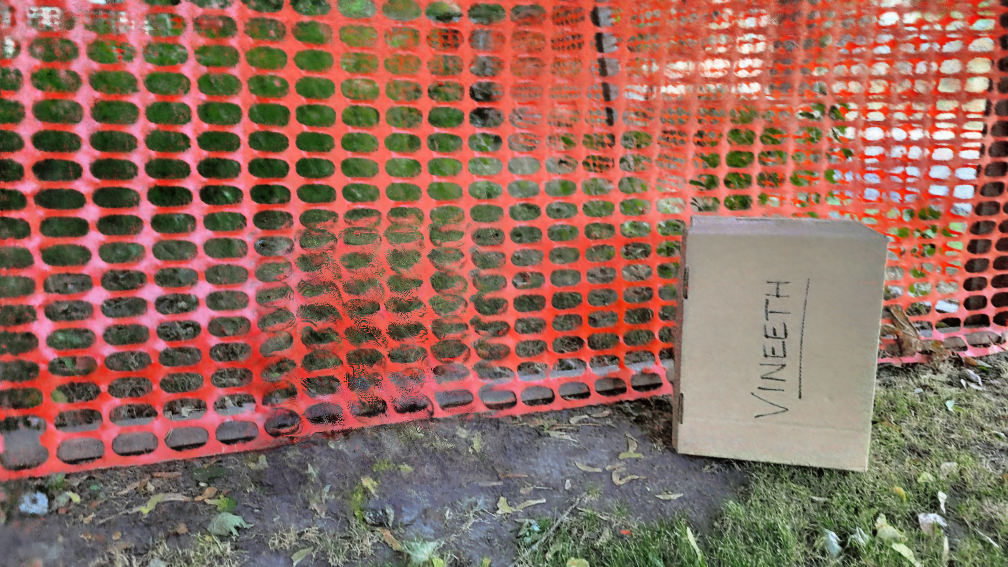} &
\includegraphics[width=0.15\linewidth]{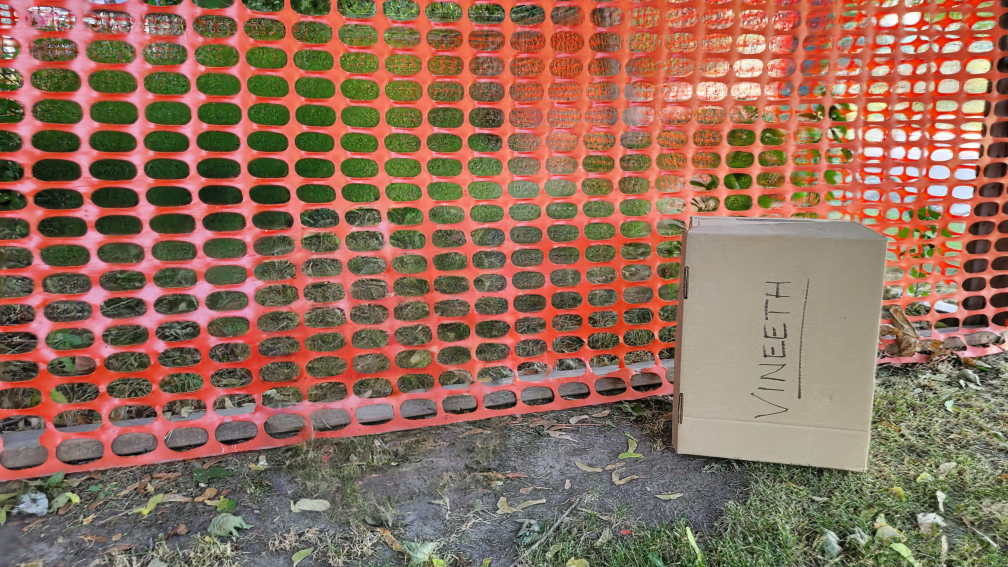} &
\includegraphics[width=0.15\linewidth]{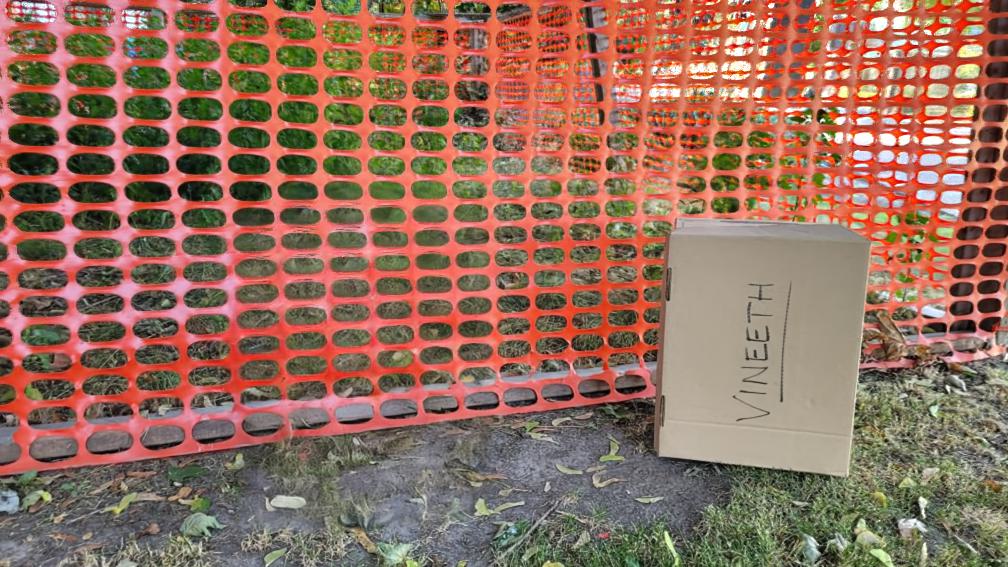} &
\rotatebox{90}{{View 1}} \\
\parbox[t]{0.15\linewidth}{\vspace{-7.0ex}\centering{(c) Scene 3}} &
\includegraphics[width=0.15\linewidth]{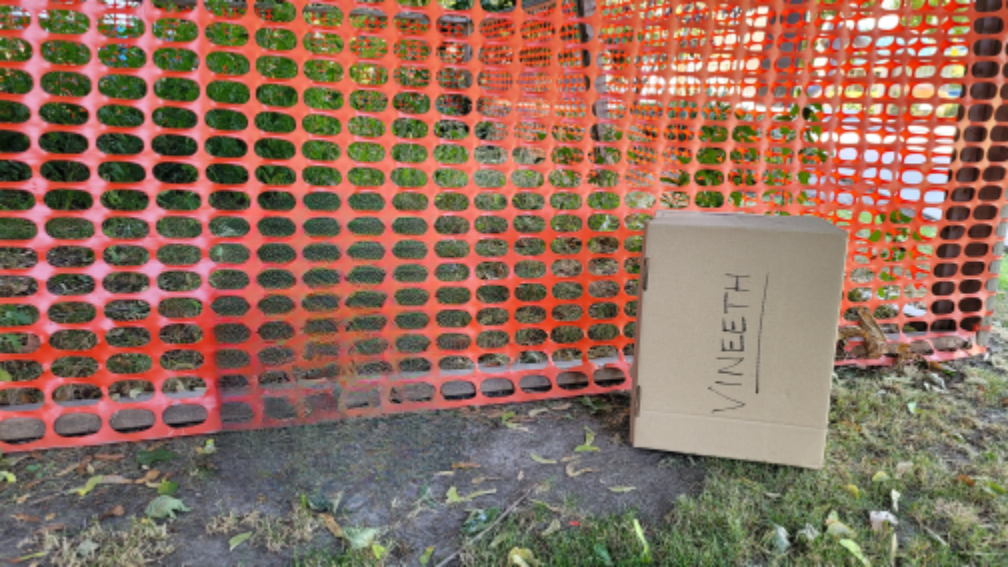} &
\includegraphics[width=0.15\linewidth]{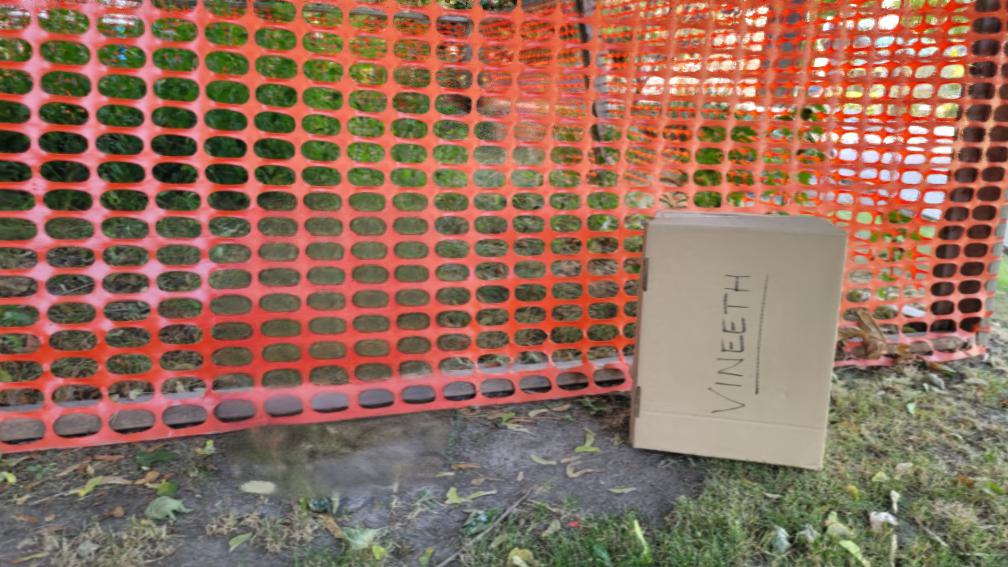} &
\includegraphics[width=0.15\linewidth]{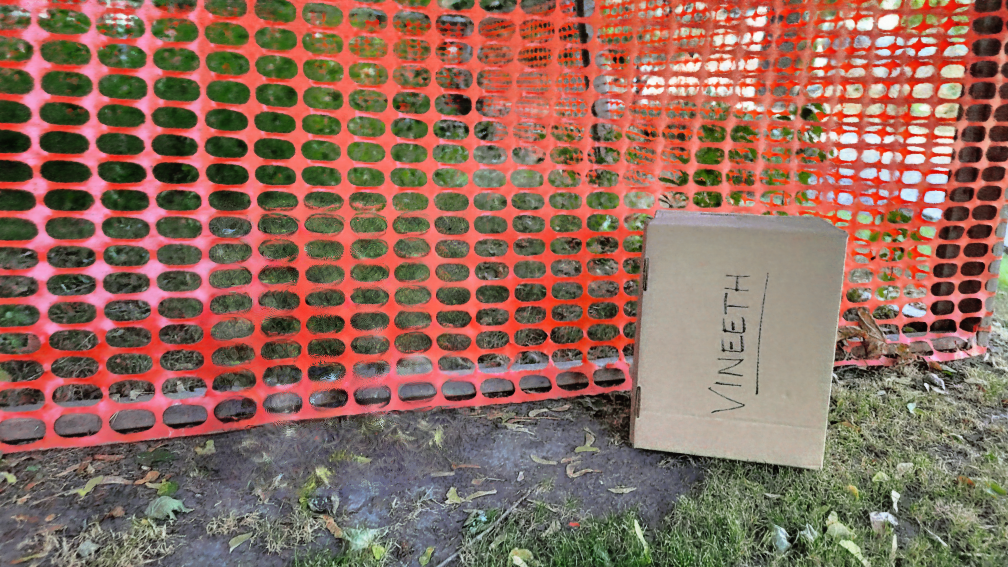} &
\includegraphics[width=0.15\linewidth]{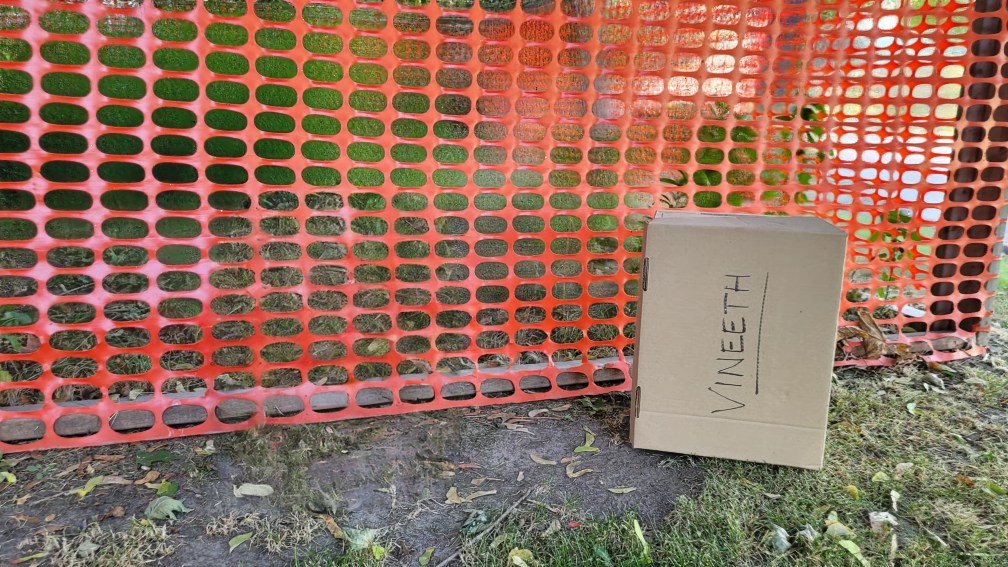} &
\includegraphics[width=0.15\linewidth]{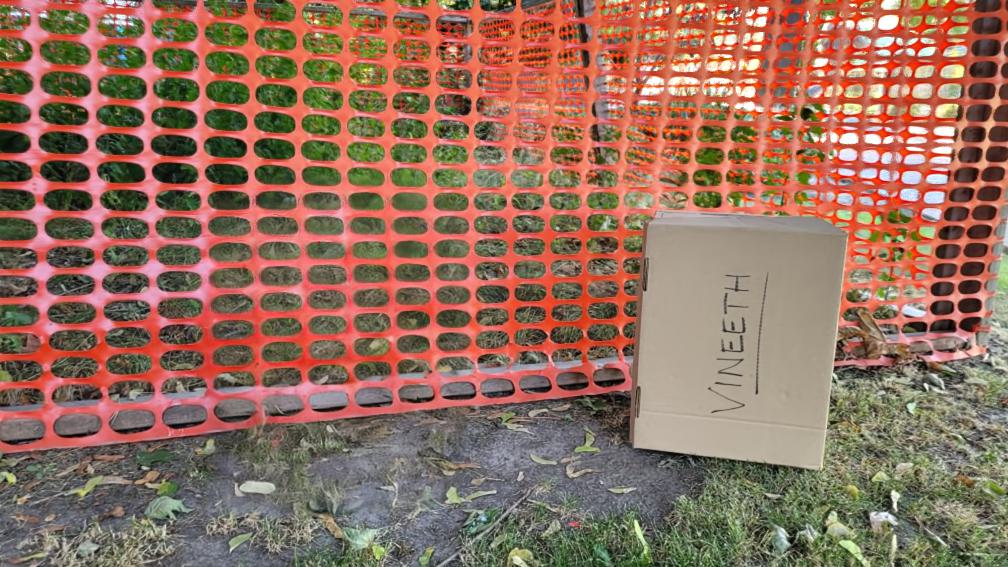} &
\rotatebox{90}{{View 2}} \\
\end{tabular}

\caption{Qualitative comparison across different methods. Each scene (a)–(c) shows inpainting results from two different viewpoints, highlighting both appearance quality and multi-view consistency.}
\label{fig:qualitative}
\end{figure}

We present quantitative and qualitative comparisons between our method and three baseline methods
in Tab. \ref{tab:quantitative} and Fig. \ref{fig:qualitative}, respectively.

\textbf{Quantitative Result} As shown in Tab.~\ref{tab:quantitative}, our method consistently outperforms GScream~\cite{wang2024gscream}, SPIn-NeRF~\cite{spinnerf}, and other baselines across all evaluation metrics, including PSNR, LPIPS and LoFTR. In particular, our approach achieves higher similarity in visual quality, as reflected by LPIPS scores, indicating that the rendered outputs more closely resemble the object-removed ground truth. Notably, unlike In-and-Out, SPIn-NeRF, and LaMask, which incorporate LPIPS-based loss functions during training, our method does not rely on such perceptual guidance, yet still delivers stronger perceptual fidelity, underscoring our approach's robustness. Furthermore, our method demonstrates stronger multi-view structural consistency, as evidenced by higher LoFTR scores across all evaluated baselines. Additionally, benefiting from the efficiency of the 3D Gaussian Splatting framework and the simplicity of our loss design, our method achieves significantly faster optimization, approximately $3 \times$ faster than GScream, offering substantial advantages in training time and computational resources. Note that MVInpainter is excluded from the training time comparison table, as it primarily operates as an inpainting module without a defined reconstruction phase—its total runtime can vary significantly depending on the choice of backend.

\textbf{Qualitative Results} in Fig.~\ref{fig:qualitative} present a visual comparison across different methods. The two most important factors affecting overall quality are color and structural semantic consistency. In scenes that require inpainting multiple textures, as seen in the second and third scenes, SPIn-NeRF tends to produce more interpolated results, although it performs very well when the inpainting region contains a single dominant texture, such as the cement surface in the first scene. MVInpainter demonstrates strong inpainting quality across various scenes; however, the 3DGS-based reconstruction stage may introduce artifacts in complex structural areas, as seen in the first and second scene. This can be attributed to the characteristics of the Stable Diffusion model used for inpainting, which preserves semantic features well but may not maintain strict pixel-level alignment required by Gaussian Splatting. In-and-Out addresses some of these reconstruction challenges by using NeRF as the 3D representation, but its results can still be impacted when large variations exist between inpainted anchor views, as observed in the second scene. Our method and GScream exhibit comparable visual quality overall; however, due to GScream’s single-anchor inpainting approach, regions farther from the anchor view can show artifacts where there is limited information, while our method maintains more consistent quality across varying poses throughout the experiments.
\subsection{Ablation Studies}
\begin{figure}
\centering
\setlength{\tabcolsep}{2pt} 
\renewcommand{\arraystretch}{1.0}

\begin{tabular}{cccc}
\multicolumn{2}{c}{Ours w/o $\mathcal{L}_{\text{SI}}
$} & \multicolumn{2}{c}{Ours} \\
\includegraphics[width=0.24\linewidth]{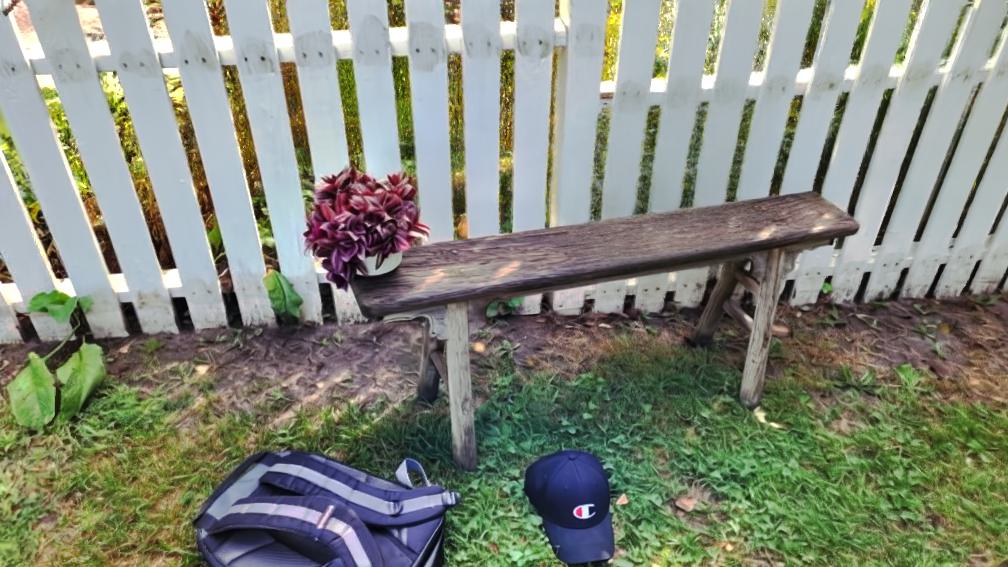} &
\includegraphics[width=0.24\linewidth]{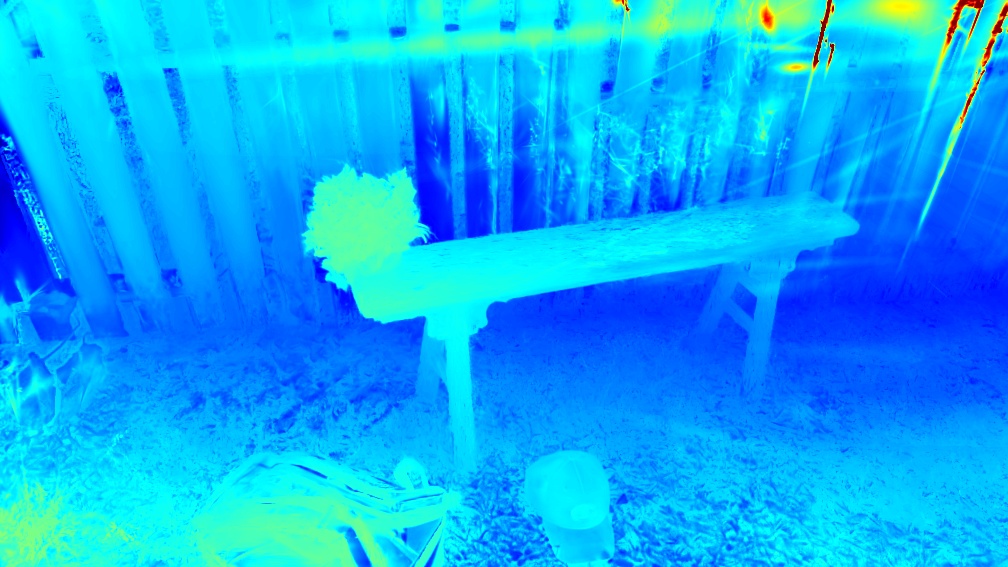} &
\includegraphics[width=0.24\linewidth]{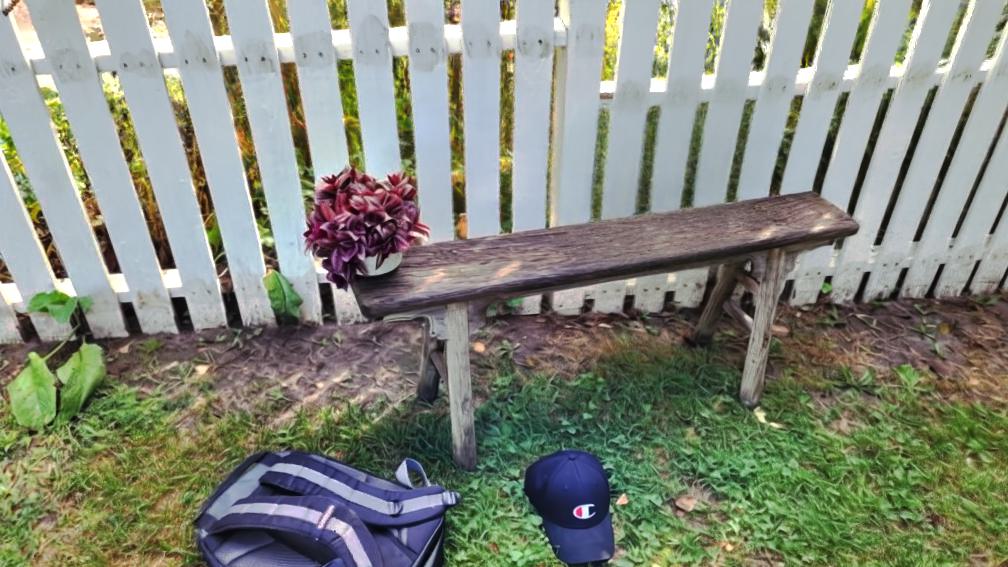} &
\includegraphics[width=0.24\linewidth]{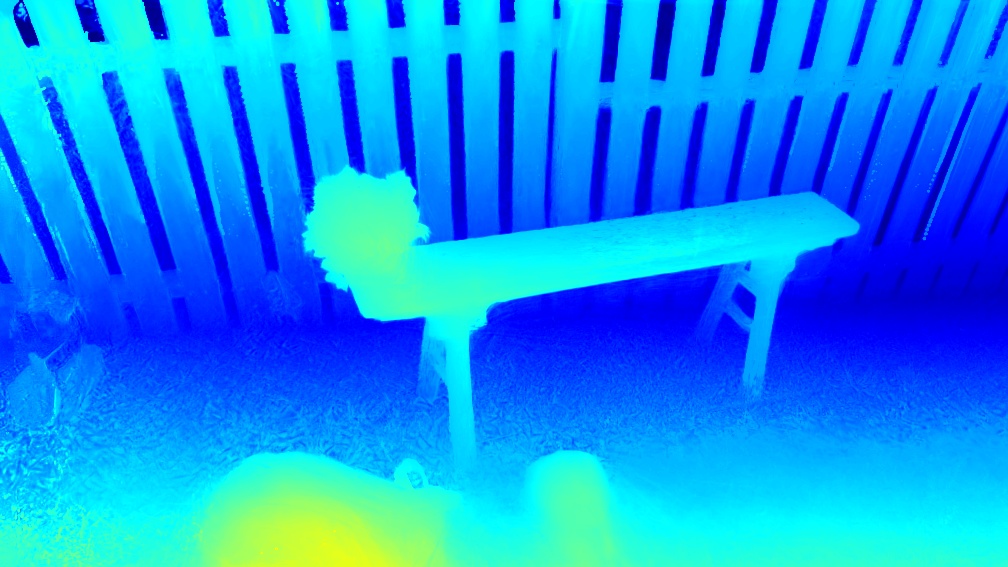} \\

\includegraphics[width=0.24\linewidth]{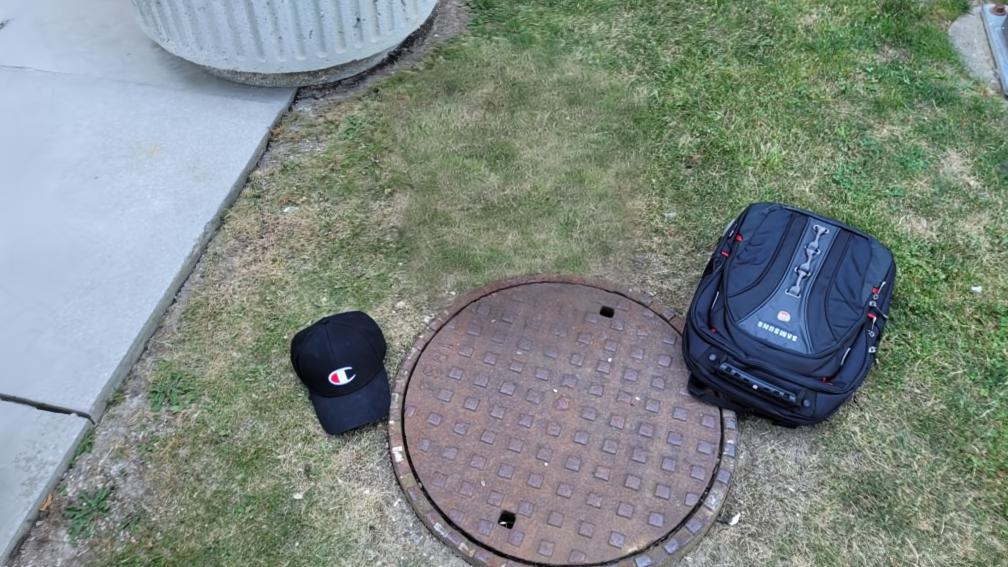} &
\includegraphics[width=0.24\linewidth]{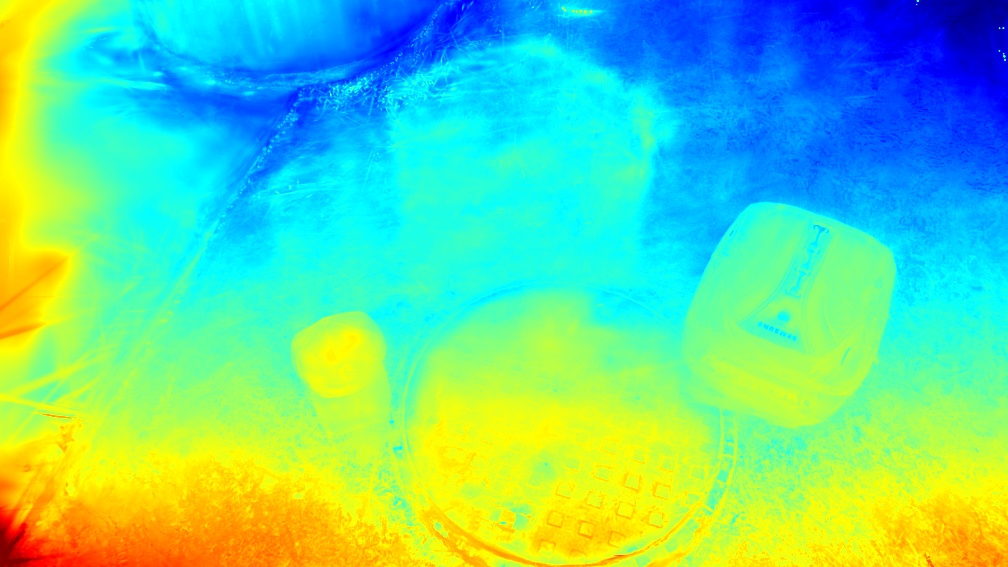} &
\includegraphics[width=0.24\linewidth]{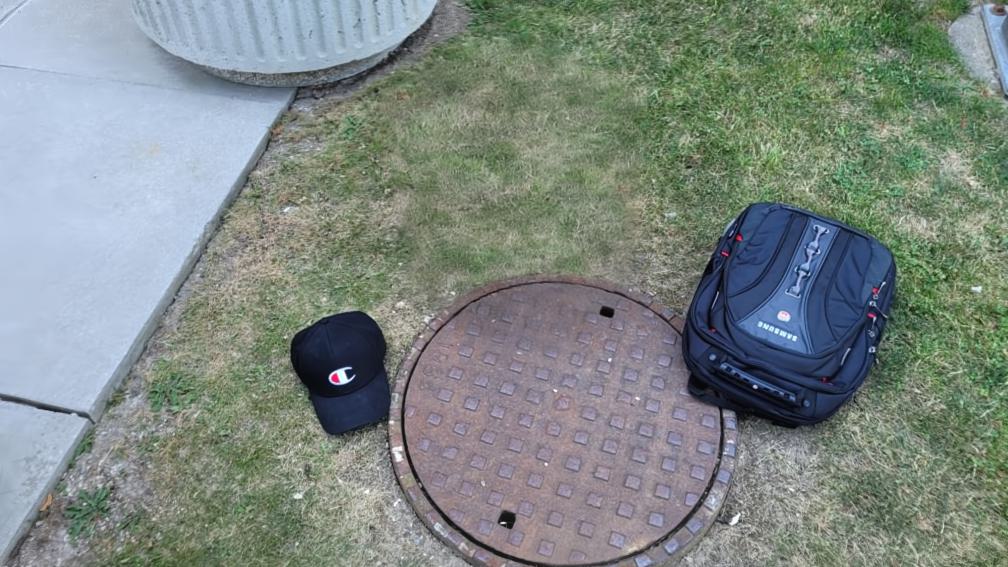} &
\includegraphics[width=0.24\linewidth]{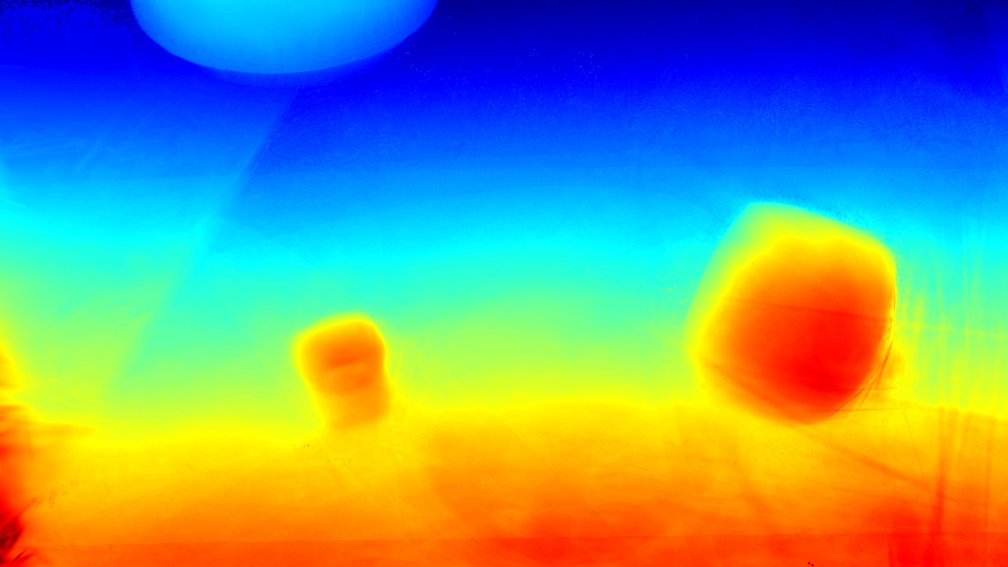} \\

\includegraphics[width=0.24\linewidth]{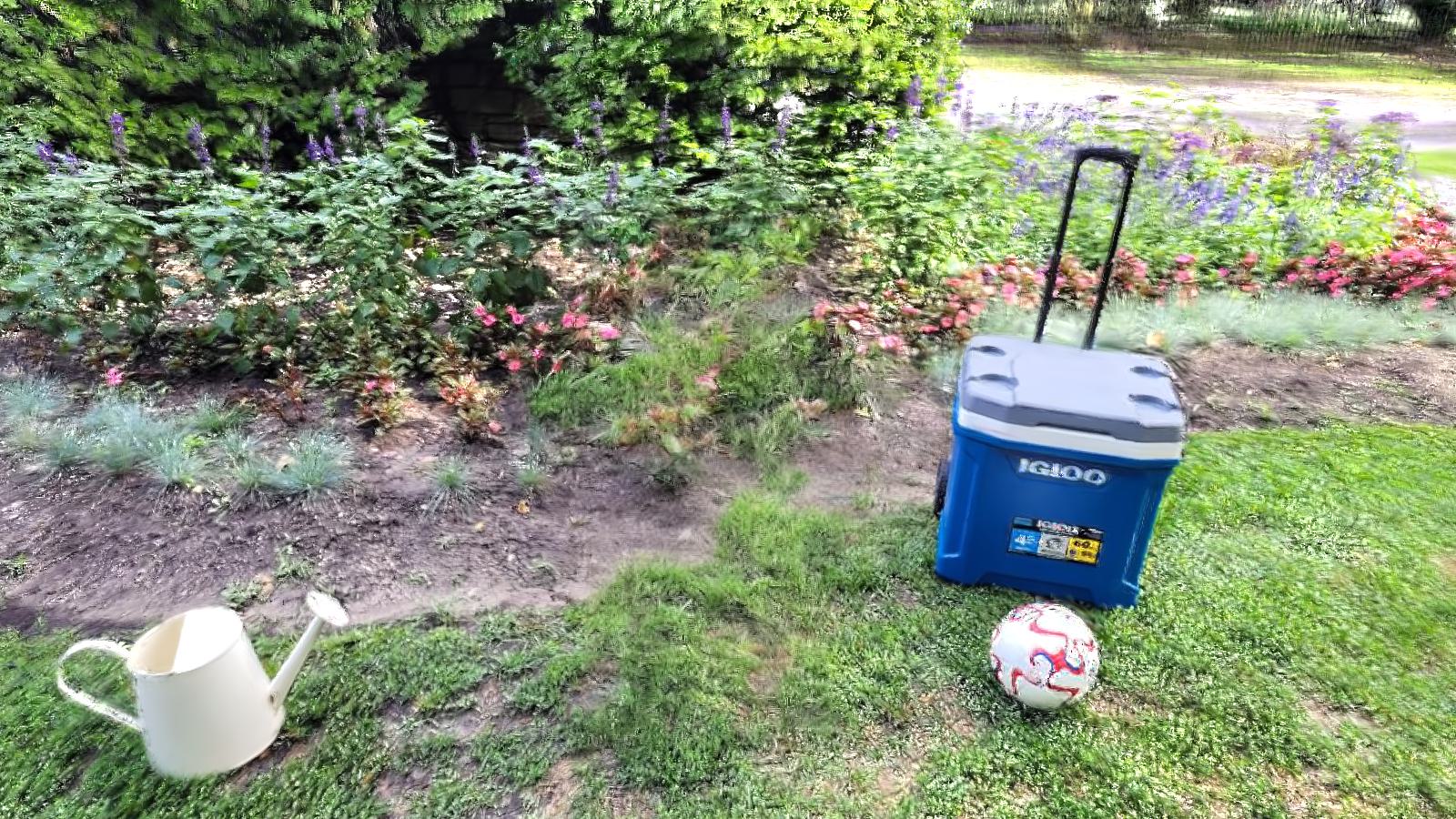} &
\includegraphics[width=0.24\linewidth]{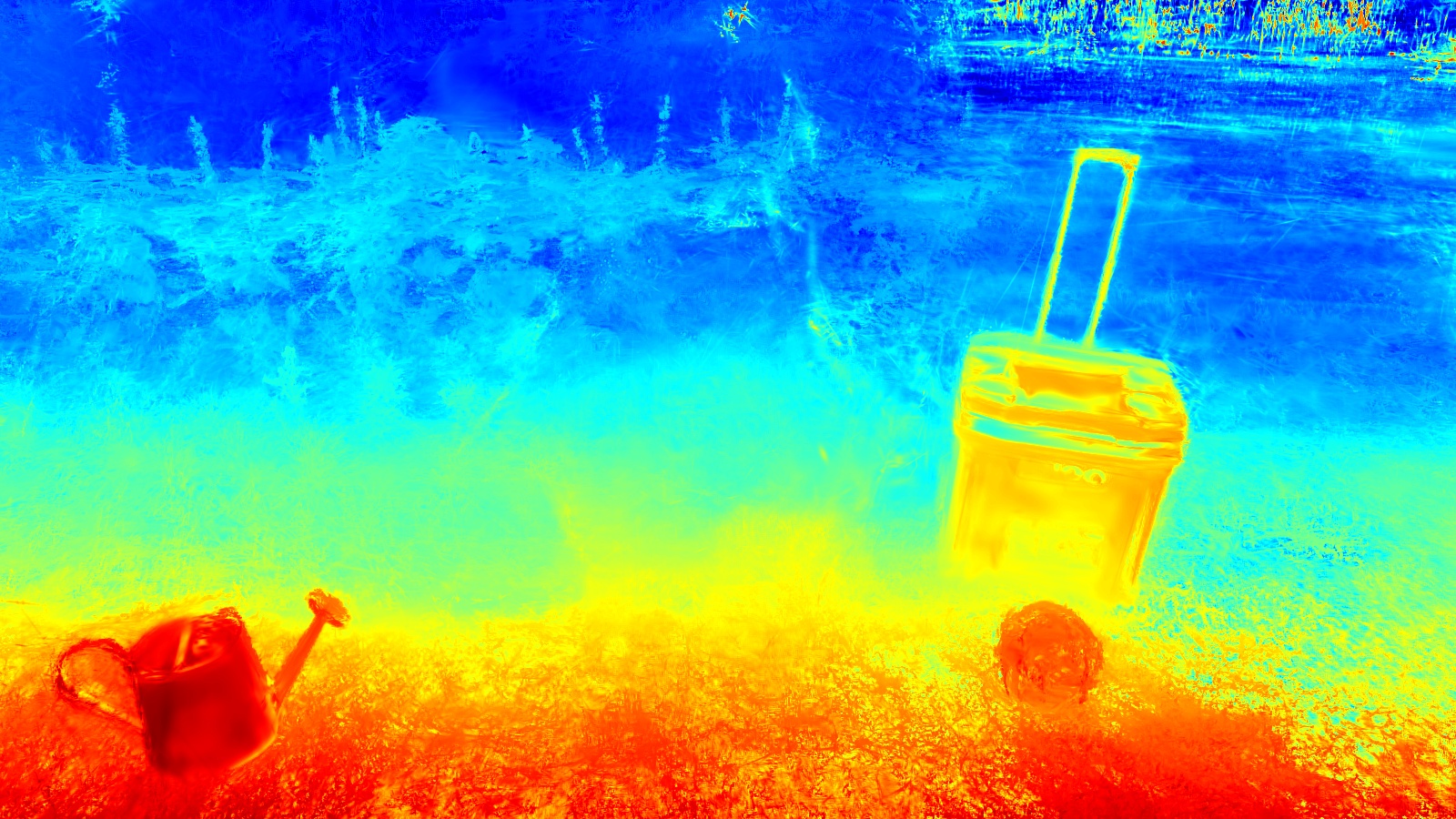} &
\includegraphics[width=0.24\linewidth]{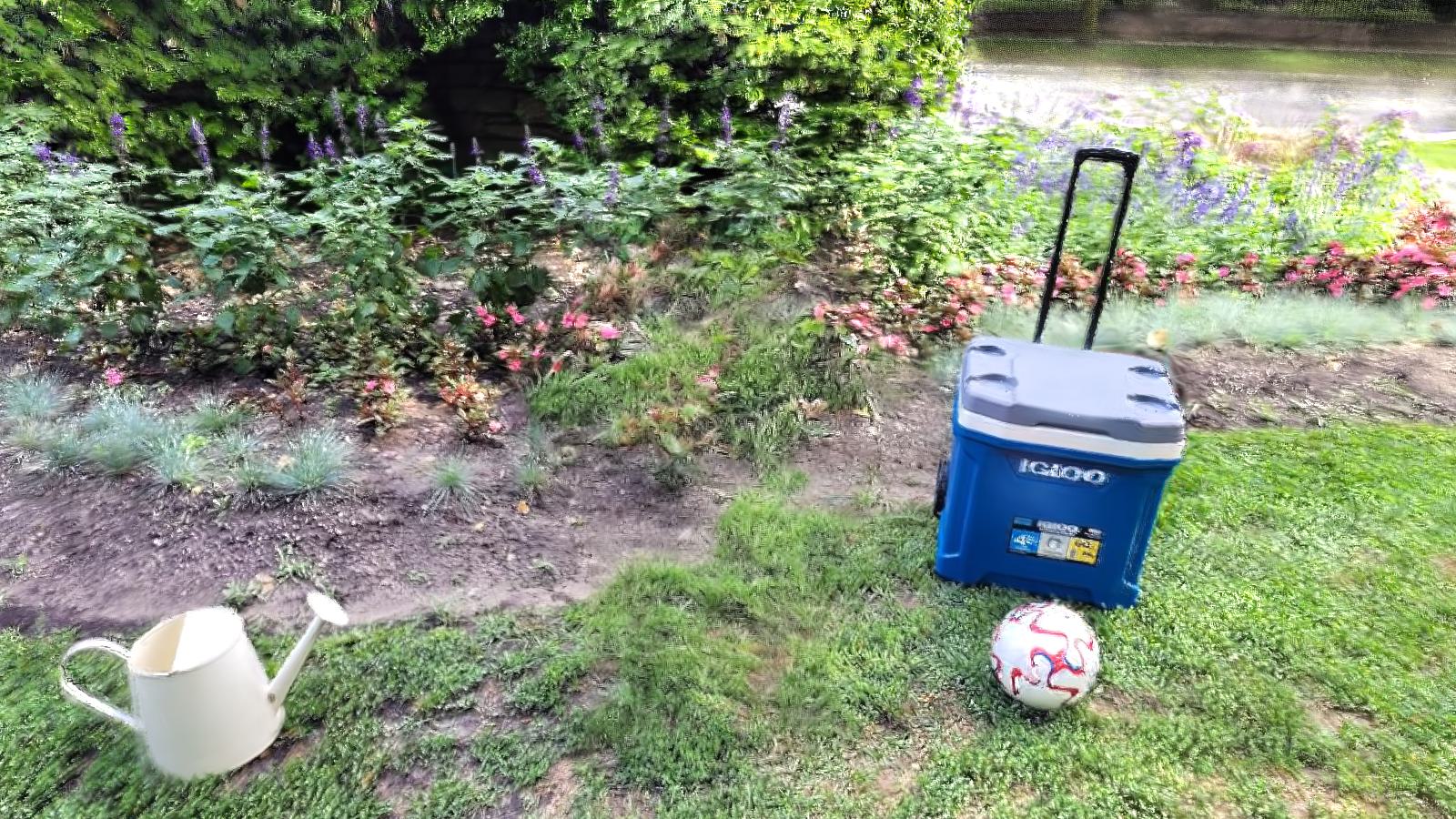} &
\includegraphics[width=0.24\linewidth]{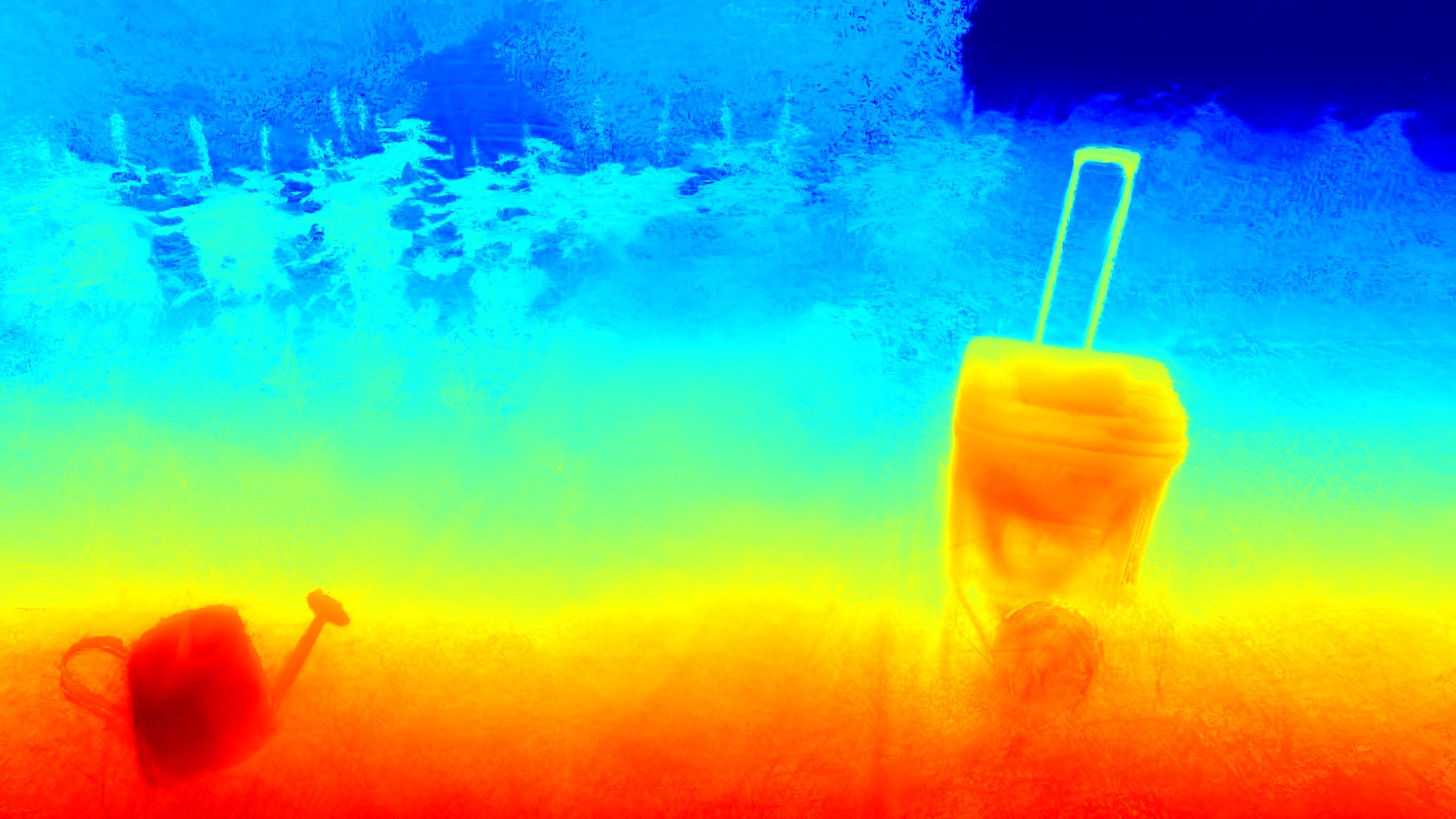} \\
\end{tabular}

\caption{Qualitative comparison between our method with and without scale-invariant depth loss. For each example, we show the inpainted RGB image and the corresponding depth map.}
\label{fig:depth-qualitative}
\end{figure}

We validate the effectiveness of the scale-invariant depth loss through ablation studies on monocular depth supervision during the reconstruction phase. Quantitative and qualitative results are presented in Table~\ref{tab:depth-ablation} and Figure~\ref{fig:depth-qualitative}.

\paragraph{Analysis of Scale-Invariant Depth Loss}
To assess the effectiveness and generality of our scale-invariant depth loss, we conduct ablations on both our method and the MVInpainter baseline. As shown in Tab.~\ref{tab:depth-ablation}, incorporating depth supervision significantly improves perceptual quality (lower LPIPS) and geometric consistency (higher LoFTR) in both pipelines. PSNR shows slight variations but remains generally stable. Meanwhile, MVInpainter, which originally lacks any form of depth supervision, also benefits from the loss formulation across all metrics, demonstrating the depth loss's general applicability. The proposed loss provides an efficient and flexible solution for enhancing 3D reconstruction quality across diverse methods and datasets.

Fig.~\ref{fig:depth-qualitative} illustrates the qualitative effects of our depth loss. Without depth guidance, inpainted regions often suffer from artifacts like semi-transparent Gaussian blobs or geometric inconsistencies, especially around occlusion boundaries (first and second rows). Applying our depth loss resolves these artifacts, producing sharper geometry and more coherent depth across views. Additionally, our method mitigates scene-wide distortions caused by input noise such as camera pose errors or imperfect initial matching (third row). These results demonstrate that our depth loss not only improves reconstructions in the edited regions but also reinforces global scene consistency, leading to more robust and photorealistic 3D outputs.

\begin{table}
\centering
\caption{Ablation results on scale-invariant depth loss for both MV-Inpainter and our method.}
\begin{tabular}{lcccc}
\hline
\textbf{Method} & \textbf{PSNR $\uparrow$} & \textbf{LPIPS $\downarrow$} & \textbf{LoFTR $\uparrow$} \\
\hline
MV-Inpainter (w/o $\mathcal{L}_{\text{SI}}
$) & 16.179 & 0.657 & 73.144 \\
MV-Inpainter & \textbf{16.320} & \textbf{0.638} & \textbf{87.262} \\
\hline
Ours (w/o $\mathcal{L}_{\text{SI}}
$) & \textbf{16.444} & 0.546 & 205.990 \\
Ours & {16.385} & \textbf{0.524} & \textbf{222.428} \\
\hline
\end{tabular}
\label{tab:depth-ablation}
\end{table}

\section{Conclusion}
\label{sec:conclusion}
We presented \textbf{VEIGAR}, a 3D scene editing framework that achieves high-fidelity and geometrically consistent reconstruction by explicitly aligning content in pixel space via deep stereo depth projection. Our scale-invariant depth loss enforces robust cross-view consistency without requiring metric depth or view-specific calibration. Experiments show that VEIGAR delivers superior visual quality and $3\times$ faster training than prior methods. The advancements enabled by VEIGAR contribute meaningfully to broader societal applications by improving the editability and accessibility of radiance fields. However, it also carries risks, as biases in the training data of diffusion models may be unintentionally reinforced from the inpainted reference during editing.

While effective, VEIGAR's performance is (1) influenced by the behavior of the inpainting model, which may interpret masks inconsistently or introduce global changes that impact alignment and visual coherence. Additionally, (2) a single reference image may be insufficient for projection in complex scenes, which can be mitigated through an iterative projection process. Despite these challenges, VEIGAR remains a practical and generalizable solution for efficient 3D editing.

\bibliography{references}

\end{document}